\documentclass[12pt, letterpaper]{article}
\usepackage[margin=2cm]{geometry}
\usepackage{changes}
\definechangesauthor[name={Pedro Pessoa},color=red]{PP}
\definechangesauthor[name={Andres Martinez},color=orange]{AM}

\usepackage{amssymb}
\usepackage{amsmath}
\usepackage{graphicx}

\usepackage{mathtools}

\usepackage{cite}

\usepackage{nicefrac}   
\newcommand{\nfrac}[2]{\nicefrac{#1}{#2}}

\usepackage{color}

\usepackage[ruled,vlined]{algorithm2e}

\newcommand{\Specs}[1]{{#1}}

\vspace{-3cm}
\title{
Inherited or produced? Inferring protein production kinetics when protein counts are shaped by a cell's division history
}

\author{Pedro Pessoa$^{1,2}$,  Juan Andres Martinez$^3$, Vincent Vandenbroucke$^3$, \\
Frank Delvigne$^3$, Steve Press\'e$^{1,2,4}$ \\ 
$^1$Center for Biological Physics, Arizona State University,
Tempe, AZ, USA\\
$^2$Department of Physics, Arizona State University,
Tempe, AZ, USA\\
$^3$Terra Research and Teaching Centre, \\ Microbial Processes and Interactions (MiPI), \\ Gembloux Agro-Bio Tech, University of Liège, Gembloux, Belgium \\
$^4$School of Molecular Sciences, Arizona State University,
Tempe, AZ, USA}
\date{}

\begin{document}

\maketitle

\abstract{
Inferring protein production kinetics  for dividing cells is complicated due to protein inheritance from the mother cell. For instance, fluorescence measurements -- commonly used to assess gene activation -- may reflect not only newly produced proteins but also those inherited through successive cell divisions. In such cases, observed protein levels in any given cell are shaped by its division history.
As a case study, we examine activation of the \emph{glc3} gene in yeast involved in glycogen synthesis and expressed under nutrient-limiting conditions. We monitor this activity using snapshot fluorescence measurements via flow cytometry, where GFP expression reflects \emph{glc3} promoter activity. A naïve analysis of flow cytometry data ignoring cell division suggests many cells are active with low expression.  
Explicitly accounting for the (non-Markovian) effects of cell division and protein inheritance makes it impossible to write down a tractable likelihood -- a key ingredient in physics-inspired inference, defining the probability of observing data given a model. The dependence on a cell's division history breaks the assumptions of standard (Markovian) master equations, rendering traditional likelihood-based approaches inapplicable. Instead, we adapt conditional normalizing flows (a class of neural network models designed to learn probability distributions) to approximate otherwise intractable likelihoods from simulated data. In doing so, we find that \emph{glc3} is mostly inactive under stress, showing that while cells occasionally activate the gene, expression is brief and transient.
}


\newpage

\section{Introduction}

Here we envision an experiment monitoring cells at various points in time (snapshot data). From this snapshot data, our goal is to estimate the production of a particular (labeled) protein species through fluorescence. 
It is common to model the dynamics of this protein's number using memoryless (Markovian) dynamics \cite{Friedman06,Shahrezaei08,Pendar13,Kumar14,Chen22,Bingjie24,Jia17,Jiao2024}. For this iteration of the problem, inferring parameters such as protein production or degradation rates is straughforward and relies on our ability to formulate a likelihood \cite{Presse23}. 
Indeed, learning parameters from data, usually termed ``inverse modeling'', generically requires one to formulate likelihoods --- probability densities for observing data sequences given parameter sets ($\theta$), $p(\text{data}|\theta)$.  Using such likelihoods, parameters are then estimated: either by finding those parameters making the sequence of the likelihood largest (from which derives the notion of \textit{maximum likelihood}); 
or by constructing posteriors over the parameters using the likelihood \cite{Bryan20,Bryan22,Presse23,Bryan23,Pessoa23}.

However, when experiments span multiple cell division cycles, a fundamental modeling challenge occurs as many proteins have half-lives longer than a single division \cite{Li00,Eden11,RusilowiczJones22}. As such, reduction in protein numbers occurs, mainly, via cell division rather than degradation. Protein counts in any given cell thus depend on the cell's division history. Put differently, the Markovian assumption--inherent to existing modeling strategies--fails as, to our knowledge, no established inference framework allows us to write down  
likelihoods in such a general setting.
Since, cell division times are not exponentially distributed \cite{Sauls16,LeTreut2021}, the cell cycle cannot be modeled using Markovian dynamics. 
Yet, despite the challenges in writing down a likelihood, it is possible to simulate protein levels under cell division \cite{Jia21,Kings23,Grima23}.

For this reason, we propose an alternative paradigm to infer kinetics of protein production in the presence of cell division: the ability to perform inference assuming we can simulate processes but cannot write down the likelihood \cite{Sukys22,Covino23,Covino25}.

Indeed, in the natural sciences, we learn to simulate dynamical models, whether deterministic or stochastic, in order to replicate  experimental output and eventually predict phenomena. These models typically depend on parameters (\emph{e.g.}, forces or velocity fields) that can be tuned to approximate the experimental output we wish to simulate. 
Simulations like these have provided significant insights across scientific areas: from reactions in solution \cite{Gillespie77,Schnoerr17}, to gene expression \cite{SzavitsNossan24,Kilic23}, neuronal activity \cite{Kriegeskorte18,Blundell18}, dynamics of infection \cite{Jalali21,Waites2021}, and even toward explaining processes involved in scientific collaboration and publishing \cite{Pessoa23,seselja22}. 

Yet, deriving likelihoods analytically is often significantly more difficult than running forward simulations. For example, while simulating chemical reactions can be straightforward via Gillespie's algorithm \cite{Gillespie77,Masuda20,Presse23}, writing down the associated likelihood involves solving the chemical master equation \cite{Munsky06,Kilic23,Pessoa24}, requiring enumerating all possible system states feasibly visited. As the number of molecular species grows, the state space increases combinatorially, making the likelihood intractable due to large matrix exponentials \cite{Moler03,Munsky06,Pessoa24}. Even this is an optimistic scenario: it applies to cases for which likelihoods can be formulated albeit at high computational cost.

For the case where they cannot even be formulated, one idea to bypass likelihood calculation using (forward) simulations to approximate the likelihood, as the simulated results can be understood as samples from $p(\text{data}|\theta)$. This leads to what is known as simulation-based inference (SBI) \cite{Covino23,Sukys22,Covino25}. An early implementation of SBI is approximate Bayesian computation (ABC), where the likelihood is approximated by the fraction of simulated data points falling within a threshold distance, $\epsilon$, of observed data \cite{Cranmer20}. However, ABC presents a number of limitation true to some ABC implementations: (1) it requires new simulations for each parameter values $\theta$, making likelihood evaluation computationally intensive; (2) it depends on arbitrary distance metrics; and (3) likelihood estimates can vary unpredictably, particularly for rare events or large datasets. 
 
To avoid these limitations, recent advances in neural networks have inspired new approaches for SBI.
Some of these consist of finding a neural network approximation of the Bayesian posterior directly, bypassing the likelihood altogether  \cite{Sierra24,Dax23,Greenberg19,Papamakarios16}. 
However, without an explicit likelihood function, we lose a key tool in validating the inferred posterior. When a likelihood is available, we can compare the approximated likelihood with histograms obtained from simulations in regions of high posterior.  In some particular benchmarks \cite{Hermans22,Falkiewicz23}, it has been argued that posterior approximations can produce overconfident posteriors underestimating uncertainty. Without an estimated likelihood to compare against, problems like these can go unnoticed. As a result, purely posterior-focused approaches have faced criticism for limited generalizability and the lack of reliable validation strategies \cite{Fengler21,Hermans22,Falkiewicz23}.

\begin{figure}
    \centering \includegraphics[width=0.9\linewidth]{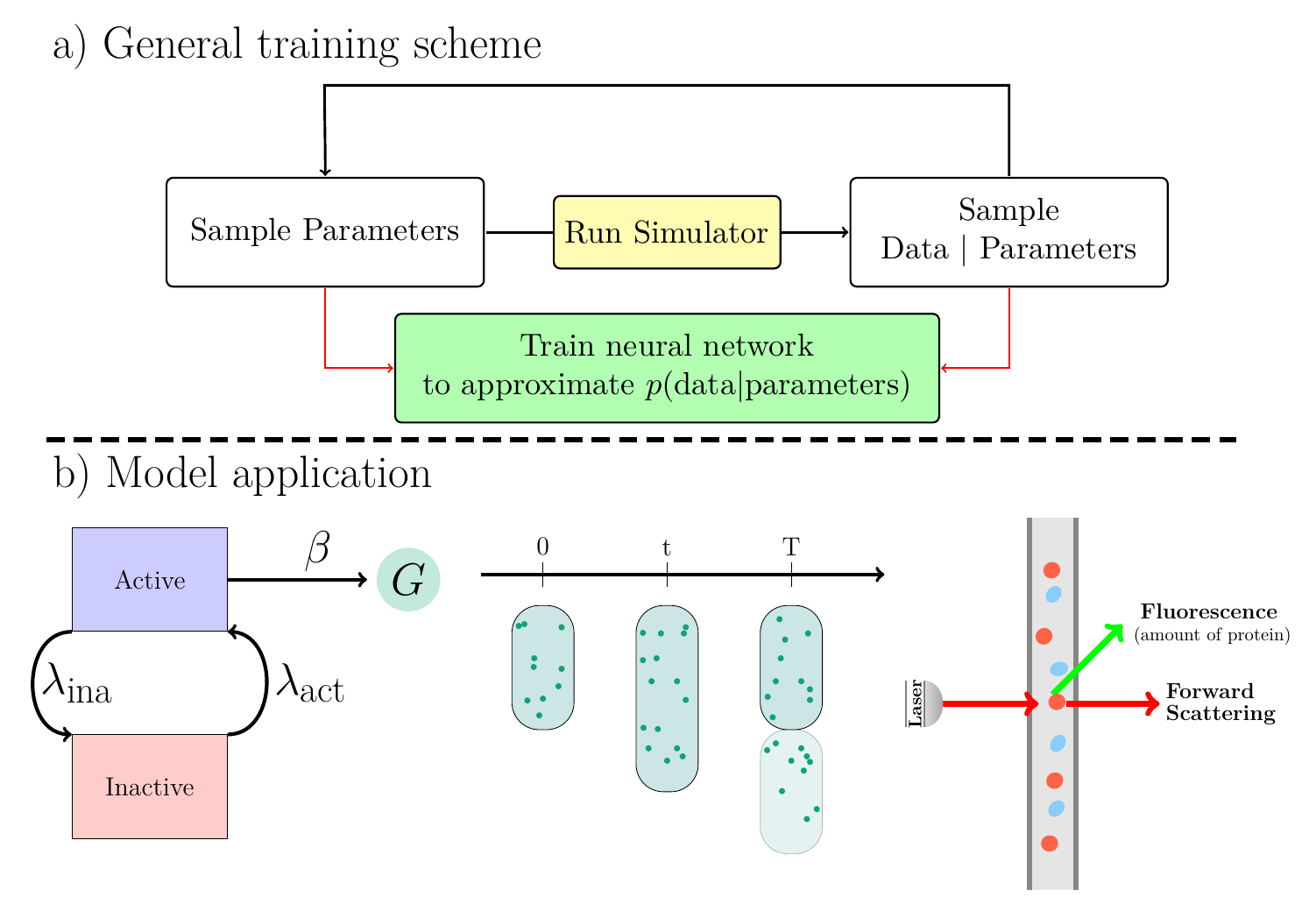}
    \vspace{-.5cm}
    \caption{
    \textbf{Summary of the simulation based inference with likelihood approximation for learning protein count dynamics through flow cytometry.}
    a) Our approach involves developing a system continuously sampling parameters and simulating the dynamics, encompassing the processes underlying both the system's behavior and the measurement process, to predict the data expected from these parameters. At each new sampling step, we train a neural network to estimate the conditional probability of the observed data given the sampled parameters.
    b) We aim to study specific systems for application, culminating in a simulator that integrates three processes: 1) protein production, including transitions driven by phenotypical state switches \cite{Martinez22}; 2) the splitting of proteins during cell division, a non-Markovian process that renders conventional techniques like solving the chemical master equation insufficient; and 3) the measurement process where, for this example, flow cytometry is employed to indirectly estimate the amount of a specific fluorescent protein per cell by measuring the total fluorescence intensity.
    }
    \label{fig:intro}
\end{figure}

In contrast, a likelihood-focused approach provides both robustness and interpretability. For each model studied, we iteratively sample parameter values, simulate corresponding datasets, and train a neural network to map parameters to the likelihood of the data. This cycle of sampling, simulation, and refinement continues until the network accurately captures the likelihood function across the parameter space (see Fig.  \ref{fig:intro}a).

Multiple neural network–based approaches have been proposed that output probability distributions and can be used to approximate likelihoods. 
For example, mixture density networks output the parameters of a predefined mixture distribution conditioned on model parameters \cite{Sukys22,Covino23}. Energy-based models define an unnormalized probability landscape via a learned energy function and typically rely on sampling techniques to estimate likelihoods \cite{Xiao24,Du24}. More recently, score-based diffusion models have been introduced, which approximate distributions by progressively corrupting data with noise and learning to reverse this process through denoising dynamics \cite{Song21,Ghio24}. Other approaches shift focus from the distribution itself to learning the underlying time evolution equations, such as the chemical master equation, using neural networks thereby replacing traditional solvers during inference \cite{Jiang21,Liu24,Nieves24}.

Here, we focus on normalizing flows due to their natural probabilistic interpretation, expressiveness, and ease of use for lower-dimensional applications considered herein. 
Normalizing flows operate by transforming a simple base distribution, typically Gaussian, through a sequence of invertible and differentiable mappings, producing a flexible and expressive target distribution.
These transformations are parameterized by neural networks and trained directly on data samples to approximate unknown densities. 
Conditional normalizing flows \cite{Papamakarios16,Trippe18,Papamakarios19,Winkler23,Zhai25} extend this framework to learn probability distributions conditioned on external variables (in our case, the model parameters $\theta$). 
For likelihood approximation, this means training the flow to approximate the conditional density $p(\text{data}|\theta)$ across the parameter space, providing an efficient and tractable surrogate for the true likelihood.

While conditional flows have been proposed for general density estimation tasks \cite{Papamakarios16,Papamakarios19}, their application as intractable likelihood approximators remains relatively underexplored in scientific modeling with recent work concentrated on classification problems \cite{Winkler23}. In this work, we broaden their scope by applying them to more general and interpretable scientific models and, in doing so, uncover the kinetics of gene state switching in yeast (\emph{S. cerevisiae}) by accurately taking into account proteins counts across cell division.

To demonstrate the flexibility of our approach and its eventual applicability to inference of protein production kinetics under cell division, we apply it to a hierarchy of three increasingly realistic biological models. We begin with a simple system in which a cell produces protein of interest at a constant rate and divides deterministically after fixed time intervals which we will later define as Model 1. 
Next, in Model 2, we extend Model 1 to include stochastic cell division times. 
Finally, in Model 3 we consider a realistic experimental setting where levels of  activation of the \emph{glc3} gene in yeast, involved in
glycogen synthesis, are monitored under nutrient-limiting conditions via fluorescent flow cytometry. By considering cell division in our inference on \emph{glc3} activity, we uncover contrary to expectation that even under high stress conditions, \emph{glc3} remains inactive in nearly all cells with rapid switches to an upregulated state. For concreteness, a high level SBI workflow and description of our models are summarized in Fig. \ref{fig:intro}.

\section{Methods}\label{sec:methods}

\subsection{Simulators, latent variables and likelihoods}

Simulations play a central role across scientific disciplines by enabling researchers to explore how different mechanisms influence observable outcomes. More precisely, a simulator is a computational tool that takes model parameters, denoted by $\theta$, and uses random number generators to sample realization-specific variables, which we collectively denote by $z$. These variables, in turn, produce observed data points, denoted by $y$, that are correlated to the underlying data-generating process  ---  such as indirect measuments of a subset of $z$.

A classic example of a simulator is the Gillespie algorithm \cite{Gillespie77, Presse23}, which simulates the stochastic dynamics of chemical reaction networks. In this context, the model parameters $\theta$ represent reaction rates, while $z$ corresponds to the full trajectory of reaction events and waiting times generated by the simulator, given these rates. 
In snapshot experiments, where only the state of the system at a specific time point is measured, most of the simulated trajectory $z$ remains unobserved. We therefore refer to $z$ collectively as latent variables. In such cases, $z$ encodes the complete underlying stochastic history, while the observed data $y$ captures only the state of the system at the measurement time point.

In probabilistic terms, the relationship between model parameters and observations is described by the likelihood $p(y|\theta)$. Since simulation involves latent variables, a simulator effectively provides access to samples from the joint distribution $p(y, z|\theta)$. The likelihood of the observed data can then be expressed by marginalizing over $z$
\begin{equation}\label{eq:marginal_likelihood}
p(y|\theta) = \int dz , p(y,z|\theta) = \int dz , p(y|z,\theta) , p(z|\theta) .
\end{equation}

Even when closed-form expressions for both $p(y|z,\theta)$ and $p(z|\theta)$ are available, evaluating the integral in \eqref{eq:marginal_likelihood} is often analytically intractable. For example, in chemical kinetics models commonly simulated with the Gillespie algorithm, the marginal likelihood $p(y|\theta)$ could, in principle, be obtained by solving the chemical master equation (CME). However, solving the CME is typically computationally intensive \cite{Munsky06,Sukys22,Pessoa24}, and formulating the CME is only feasible for systems with Markovian dynamics. Moreover, when the latent space $z$ is high-dimensional, even numerical integration becomes prohibitively expensive.

Nevertheless, even when the marginal likelihood in \eqref{eq:marginal_likelihood} cannot be evaluated analytically, the simulator still allows us to generate samples from it. Provided that the simulator faithfully represents the data-generating process, drawing samples from the joint distribution $p(y, z|\theta)$ ensures that the resulting observations $y$ follow the correct marginal distribution $p(y|\theta)$, simply by ignoring $z$.

This ability to sample from $p(y|\theta)$ directly enables us to approximate the likelihood without requiring explicit evaluation. In practice, we can use these samples to train a neural network to approximate the distribution $p(y|\theta)$. In the following section, we describe how this can be achieved using normalizing flows.

\subsection{Normalizing flows for likelihood approximation}\label{sec:training_general}

While simulators provide a means to generate synthetic data for known parameters, they do not, by themselves, enable inference. The key scientific challenge is to invert the simulation process: to recover the underlying parameters $\theta$ from observed data $y$. This requires evaluating the likelihood $p(y|\theta)$ across a range of candidate parameters. 
When it is infeasible to perform multiple likelihood calculations, we are inclined to use neural networks to approximate the likelihood function, treating simulations as the effective training data: by generating many $(\theta, y)$ pairs, simulators allow us to learn an approximate likelihood directly from samples in otheer words, perform SBI with neural networks \cite{Sukys22,Covino23,Covino25}. 

A particularly powerful architecture for this task is the normalizing flow. Normalizing flows \cite{Rezende15,Papamakarios17,Durkan19,Papamakarios21,Stimper23,Du24,Pessoa25a}  are a powerful tool, providing a mechanism to evaluate likelihoods by constructing a flexible family probability distributions. They work by learning differential and bijective transformations mapping a simple base distribution into the distribution we aim to approximate.

Specifically, a normalizing flow learns a transformation $f_\phi$, parameterized by the internal parameters of the neural network, collectively denoted as $\phi$, such that $y = f_\phi(\xi)$. This transformation allows us to express the probability density of $y$ as
\begin{equation}
p_\text{NF}(y|\phi) = p_\xi(\xi) \left| \det \frac{\partial f_\phi^{-1}}{\partial y} \right|^{-1},
\end{equation}
where $\xi = f_\phi^{-1}(y)$ is the inverse transformation, $\det \partial f_\phi^{-1}/\partial y$ is the determinant of the Jacobian, and $p_\xi(\xi)$ is the base distribution, usually a standard Gaussian.

For likelihood approximation, the transformation must additionally be conditioned on the model parameters $\theta$, yielding
\begin{equation}
p_\text{NF}(y | \theta, \phi) = p_\xi(\xi) \left| \det \frac{\partial f_\phi^{-1}(y; \theta)}{\partial y} \right|^{-1}.
\end{equation}
This framework, known as conditional normalizing flows \cite{Papamakarios16,Trippe18,Papamakarios19,Winkler23}, enables the model to train the neural network to learn the conditional distribution $p(y|\theta)$. In other words, to find the parameter values $\phi^\ast$ such that
\begin{equation}\label{nf_approx}
p_\text{NF}(y | \theta, \phi^\ast) \approx p(y | \theta).
\end{equation}

To train the normalizing flow to satisfy \eqref{nf_approx}, we minimize the discrepancy between the true likelihood $p(y|\theta)$ and its approximation $p_\text{NF}(y|\theta,\phi)$. A standard way to measure this discrepancy is the Kullback-Leibler (KL) divergence \cite{Kullback51,Shore80,Vanslette17,Pessoa21} defined as
\begin{equation} \label{KL_def}
    \text{KL}(\theta,\phi) =  \int dy \, p(y|\theta) \log  \frac{p(y|\theta)}{p_\text{NF}(y|\theta, \phi)}.
\end{equation}
To learn the likelihood across a range of parameter values $\theta$, we extend the KL divergence by taking the expected value over $\theta$ from a prior distribution $p(\theta)$
\begin{equation}\label{KL_theta} 
 \text{KL}(\phi) = \int d\theta \ p(\theta) \text{KL}(\theta,\phi)  = \int d\theta \ p(\theta) \int dy \ p(y|\theta) \log \frac{p(y|\theta)}{p_\text{NF}(y|\theta, \phi)}.
\end{equation} 

However, evaluating this expression requires explicit knowledge of the true likelihood $p(y|\theta)$, which is unavailable. 
Instead, we can draw a sequence of parameter samples $\{\theta_s\} = \{\theta_1, \theta_2, \ldots, \theta_S\}$ from the an \textit{a priori} distribution,  $\theta_s \sim p(\theta)$ which will later be used as the Bayesian prior, and generate corresponding synthetic data points $\{y_s\} = \{y_1, y_2, \ldots, y_S\}$ from their respective parameters, $y_s \sim p(y|\theta_s)$, using the simulator, as depicted in Fig.\ref{fig:intro}a, this procedure generates the set of simulated pairs $(y_s,\theta_s)$.  These samples allow one to approximate the KL divergence in \eqref{KL_theta} as a Monte Carlo integration yielding 
\begin{equation}\label{KL_final}
\begin{aligned}
    \text{KL}(\theta,\phi) &\approx \frac{1}{S} \sum_{s=1}^{S} \left( \log \frac{p(y_s|\theta_s)}{p_\text{NF}(y_s|\theta_s, \phi)} \right)\\ 
    &= C -  \frac{1}{S} \sum_{s=1}^{S} \log {p_\text{NF}(y_s|\theta_s, \phi)}. 
\end{aligned}
\end{equation}
where $C$ is a constant absorbing the intractable term involving the true likelihood $p(y_s|\theta_s)$, which does not depend on the neural network parameters $\phi$ and therefore does not affect optimization.

As a result, minimizing the KL divergence in \eqref{KL_final} is equivalent to minimizing the empirical negative log-likelihood of the simulated data
\begin{equation}\label{negloglike}
\mathcal{L}(\phi) = -\frac{1}{S} \sum_{s=1}^{S} \log p_\text{NF}(y_s | \theta_s,\phi)
\end{equation}
and obtaining  
\begin{equation}
    \phi^\ast = \arg \min_\phi   \mathcal{L}(\phi).
\end{equation}
 
This approach allows us to sidestep the need to compute the true KL divergence, relying instead on evaluating the conditional normalizing flow probability $p_\text{NF}(y_s|\theta_s)$ for simulated pairs. 
As training depends on these simulated $(\theta_s, y_s)$ pairs, we adopt a dynamic resampling strategy during training to avoid overfitting to a finite number of simulations. 
As illustrated in Fig.\ref{fig:intro}a, each training cycle consists of sampling a new set of parameter values $\{\theta_s\}$ from the prior and running the simulator to generate corresponding synthetic data $\{y_s\}$. 
These newly generated pairs are then used to calculate the loss function defined in \eqref{negloglike}, treating $(\theta_s, y_s)$ as training examples. 
This cycle (sampling parameters, simulating observations, and updating $\phi$ in order to minimize \eqref{negloglike}) is repeated throughout the training to ensure that the neural network learns a smooth and generalizable approximation of the likelihood across the full parameter space. 
The specific details of the resampling frequency, dataset sizes, and replacement strategies depend on the model under study and are described in Sec.\ref{sec:allmodels}.

\subsection{Bayes' theorem and parameter learning}

Having trained the normalizing flow $p_\text{NF}(y|\theta,\phi^\ast)$ to approximate the likelihood $p(y|\theta)$, we now turn to the problem of inferring model parameters $\theta$ from real observed data. We denote the observed dataset as 
$\{ y^n \} = \{ y^1, y^2, \ldots, y^N \},$
where each $y^n$ represents an experimental measurement. This observed data should not be confused with the simulated samples used during training, denoted as $y_s$.

From Bayes' theorem, the distribution for $\theta$ given $\{y^n\}$, the posterior, is
\begin{equation}\label{bayes}
p( \theta | \{ y^n \} ) = \frac{ p( \{ y^n \} | \theta ) p( \theta ) }{ p( \{ y^n \} ) }  \propto p( \{ y^n \} | \theta ) p( \theta ).
\end{equation}
where $p(\theta)$ is the prior over the parameters, and $p(\{ y^n \})$ is the evidence, independent of $\theta$, thus allowing us to write the posterior up to proportionality as above.

Assuming that the data points are conditionally independent given $\theta$, the total likelihood factorizes as $p( \{ y^n \} | \theta ) = \prod_{n=1}^N p( y^n | \theta )$. 
We can now replace each likelihood factor with the output of the trained normalizing flow, setting $p( y^n | \theta ) \approx p_\text{NF}( y^n | \theta, \phi^\ast )$.
Substituting this into the factorized likelihood, the posterior becomes
\begin{equation}\label{eq:posterior_nf}
p( \theta | \{ y^n \} ) \propto \left( \prod_{n=1}^N p_\text{NF}( y^n | \theta, \phi^\ast ) \right) p( \theta ),
\end{equation}
{In the following sections, we apply this framework to specific models of interest. For each case, we define the forward process that generates synthetic data, train a conditional normalizing flow to approximate the likelihood, and perform Bayesian inference based on the observed experimental data. }

\section{Results}\label{sec:allmodels}

Here, we evaluate the performance of our likelihood approximation framework, in a sequence of models of increasing biological complexity:

\begin{itemize}
    \item \textbf{Model 1: Regular cell division (Sec. \ref{sec:deterministic_cell_division})}  This model admits an exact analytical expression for the steady-state likelihood serving as a reference case, and allowing us to quantitatively compare our neural network-based approximation to the exact solution.
    \item \textbf{Model 2: Stochastic cell division (Sec. \ref{sec:stochastic_cell_division})}  We introduce stochasticity in division times rendering the likelihood analytically intractable. We validate the neural network approximation by comparing the predicted likelihood at given parameters to simulation results.
    \item \textbf{Model 3: Stochastic cell division with two-state regulation and uneven partitioning (Sec. \ref{sec:fc})}  
    This model introduces additional biological complexity, including state switching regulation, asymmetric division, and indirect fluorescence measurements. It is fully simulation-based and designed to reflect realistic \emph{S. cerevisiae} protein dynamics, ultimately capable of replicating the experimental conditions recovered in \cite{Martinez22} for which we present the inference results in Sec. \ref{sec:real}.
\end{itemize}

\subsection{Model 1  ---  Regular cell division model}\label{sec:deterministic_cell_division}

The first model describes a stochastic system in which protein production occurs at a constant rate $\beta$, and cells divide at regular intervals $T$. Importantly, this model admits an exact analytical expression for the steady-state likelihood of protein counts, allowing us to directly compare the neural network-based likelihood approximation to the exact result.

The system includes the following sources of stochasticity: (1) Protein synthesis is modeled as a birth process, meaning the number of proteins produced within a time interval $\Delta t$ follows a Poisson distribution with mean $\beta \Delta t$;
(2) Cell division events are asynchronous, meaning the cell can be found in any stage of the cell cycle, independent of other cells in the environment. We denote by $\tau \in (0, T)$ as the time since the last division event. 
Each cell has a $\tau$ uniformly distributed at any point in the cell cycle, $p(\tau) = 1/T$ for $\tau \in (0, T)$;
(3) The partitioning of proteins between the two daughter cells is idealized as Binomial for this simple model. Put differently, each protein is transferred to one of the daughter cells with probability $1/2$. Thus, if the parent cell contains $G_T$ proteins preceding division, the number of proteins allocated to one daughter cell follows a Binomial distribution with $G_T$ trials and a success probability of $1/2$. 

\subsubsection{Steady-state distribution}

Here, we show that under the sources of stochasticity enumerated above (points 1–3), it is possible to derive a steady-state distribution for the number of proteins, denoted by $G$. 
The stochastic dynamics corresponding to conditions 1 and 3 can be formalized mathematically as
\begin{subequations}\label{constraints_steady}
\begin{align}
    G - G_0 | \beta, \tau &\sim \text{Poisson}(\beta \tau), \label{constraint1} \\
    G_0 | \beta &\sim \text{Binomial}(G_T, 1/2), \label{constraint2}
\end{align}
\end{subequations}
where $\sim$ denotes ``sampled from" or ``distributed as". 
Above, $G_0$ and $G_T$ refer to the protein counts immediately after and immediately before a division event, respectively.
Since we later marginalize over $\tau$, we simplify the notation by treating $G$ as the protein count at an arbitrary stage in the cell cycle, with $\tau$ being a latent variable.

A distribution consistent with the conditions in \eqref{constraints_steady} is
\begin{equation}\label{gtau}
    G | \beta, \tau \sim \text{Poisson}(\beta (1 + \tau)).
\end{equation}
From the expression above, we can recover the distributions of $G_0$ and $G_T$ by evaluating $G_\tau$ for any value of $\tau$. In particular, setting $\tau = 0$ and $\tau = 1$ yields:
\begin{subequations}
\begin{align}
    G_0 | \beta &\sim \text{Poisson}(\beta), \label{g0} \\
    G_T | \beta &\sim \text{Poisson}(2\beta). \label{gT}
\end{align}
\end{subequations}
As a sanity check, these results are consistent with the constraints in \eqref{constraints_steady}. For instance, as the sum of two independent Poisson random variables is itself Poisson distributed, with the mean equal to the sum of the individual means, the combination of \eqref{gtau} and \eqref{g0} satisfies \eqref{constraint1}.  
Second, a well-known property of Poisson distributions is that partitioning a Poisson-distributed variable via a Binomial process results in another Poisson-distributed variable, where the mean is scaled by the Binomial success probability. Thus, \eqref{g0} and \eqref{gT} are consistent with \eqref{constraint2}.

Now, in order to have a closed-form expression for $p(G | \beta)$, we observe that  the partitioning of proteins follows a Binomial between daughter cells), we observe cells at uniformly distributed times within the cell cycle, allowing us to marginalize over $\tau$ in \eqref{gtau} as
\begin{equation}\label{poisson_beta_marginalized}
\begin{aligned}
    p(G | \beta) &= \int d\tau \, p(\tau) p(G | \beta, \tau) \\
                 &= \int_0^T  d\tau  \ \frac{1}{T} \frac{e^{-\beta (1+\tau)} (\beta (1+\tau))^G}{G!} \\
                 &= \frac{\gamma(G+1, 2\beta) - \gamma(G+1, \beta)}{\beta T}.
\end{aligned}
\end{equation}
where $\gamma(a, x)$ represents the incomplete gamma function
\begin{equation}\label{igamma_def}
    \gamma(a, x) = \frac{1}{\Gamma(a)} \int_0^x  dt \ t^{a-1} e^{-t} .
\end{equation}

\begin{figure}[t]
    \centering \vspace{-.5cm}
    \includegraphics[width=0.9\linewidth]{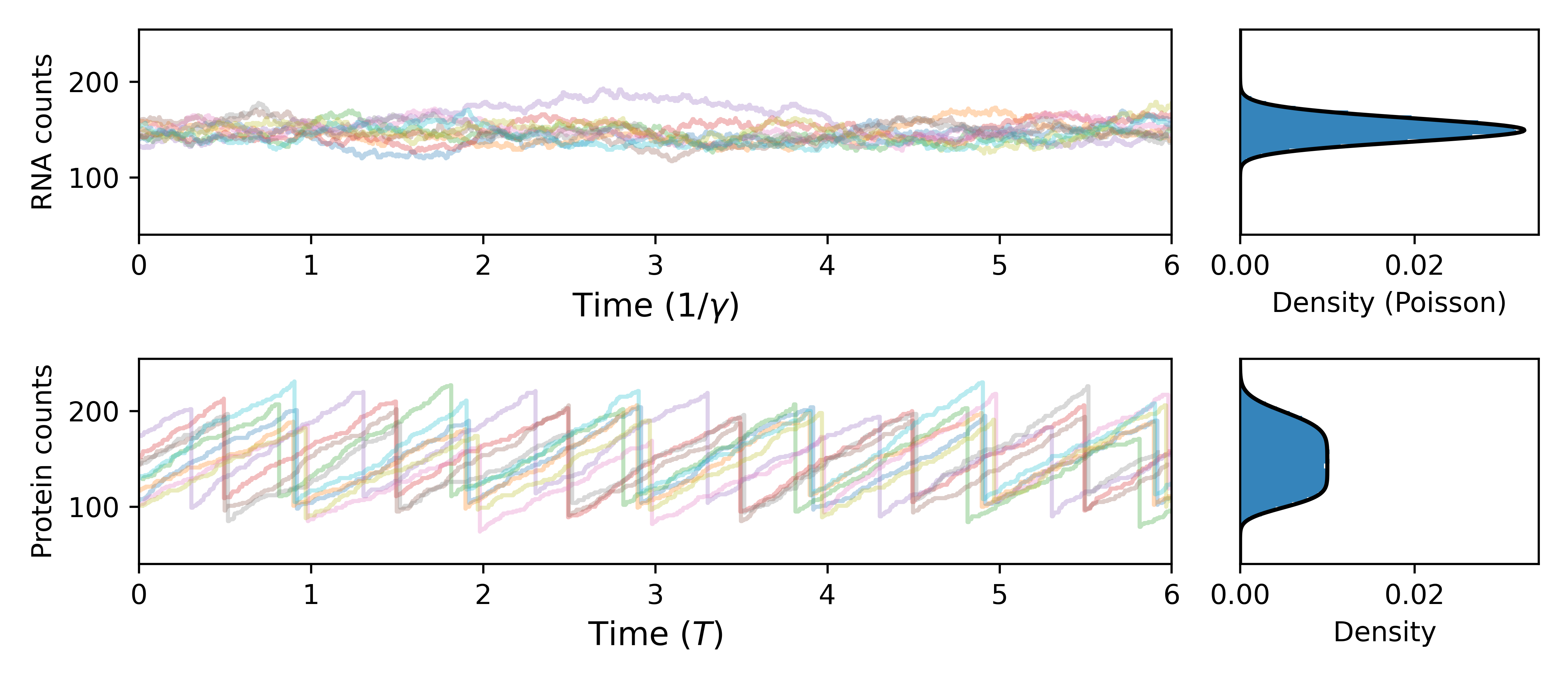} \vspace{-.7cm}
    \caption{
    \textbf{Why the Markovian approach is insufficient to learn protein production counts across cell division.}  
    RNA production is often modeled as a birth-death process \cite{Kilic23,Pessoa24} (simulated in the top left panel), where constant production leads to a Poisson steady-state distribution (top right).
    While previous literature extend these Markovian models to account for protein production\cite{Friedman06,Shahrezaei08,Pendar13,Kumar14,Chen22}, marker proteins have lifetimes on the order of hundreds of hours, thus making normal elimination by mean of  degradation negligible as compared to the reduction caused by cell division.
    In the case where protein production occurs at a constant rate and cell divisions happen deterministically at equal intervals (described in Sec. \ref{sec:deterministic_cell_division} and simulated in the bottom left), the steady-state distribution becomes an incomplete Gamma distribution \eqref{poisson_beta_marginalized} (bottom right), which is qualitatively different from the Poisson distribution recovered for birth-death process. 
    However, more complex scenarios, such as variability in division times (introduced in Sec. \ref{sec:stochastic_cell_division}) or non-constant production rates (discussed in Sec. \ref{sec:fc}) complicate the model even further, necessitating a simulation-based approach to accurately model protein counts.
    }
    \label{fig:simulated_forward}
\end{figure}

This analytical steady-state distribution was obtained from a model where proteins disappear through partitioning at cell division rather than through degradation. 
This contrasts with the classical birth–death framework commonly used in the literature \cite{Friedman06, Shahrezaei08, Pendar13, Kumar14, Chen22, Bingjie24, Jia17, Jiao2024}, where individual proteins degrade at a constant rate $\gamma$. 
Supposing a constant protein production rate $\beta$, the steady-state protein distribution in the birth–death model is Poisson with mean $\beta / \gamma$.
To highlight the dramatic qualitative differences between birth-death process versus our cell-division process, Fig. \ref{fig:simulated_forward} shows a comparison between simulations of a birth–death process with parameters $\beta = 150 \gamma$, and the cell-division process described here with $\beta = 100/T$, adjusted to achieve the same expected steady-state mean according to \eqref{poisson_beta_marginalized}.

Having established the steady-state distribution for protein counts in \eqref{poisson_beta_marginalized}, along with a qualitative comparison to the classical birth-and-death framework illustrated in Fig. \ref{fig:simulated_forward}, we have a basis to which to compare the results of our likelihood approximation scheme for this model.

\subsubsection{Training the neural network}\label{sec:training_model1}
Our goal is to train a neural network to approximate the likelihood, \textit{i.e.}, the steady-state distribution of protein counts.
Having an exact analytical solution for the steady-state distribution in \eqref{poisson_beta_marginalized}, we use the regular cell division process model as an initial test case to train and evaluate the network,  before moving on to models where the exact distribution is not available.

When training the network, we work in log space for both the parameters and the protein counts $G$. Specifically, we sample values of $\ln \beta$, transform back to $\beta$, run the simulator with $\beta$ to obtain protein counts $G$, and then take the logarithm again to obtain $\ln G$. 
Thus, $\ln G$ takes the role of the observations ($y$ in Sec. \ref{sec:methods}), while $\ln \beta$ takes the role of the parameters ($\theta$ in Sec. \ref{sec:methods}). 
Training the network consists of finding $\phi^\ast$ such that
\begin{equation}
    p(\ln G | \ln \beta) \approx p_\text{NF}(\ln G | \phi^\ast, \ln \beta).
\end{equation}
If the original distribution over $G$ is needed, it can be recovered as
\begin{equation}
\begin{aligned}
    p(G | \beta) 
    &= p(\ln G |  \ln \beta) \ \frac{d\ln G}{dG} \\
    &\approx \frac{1}{G} \ p_\text{NF}(\ln G | \phi^\ast, \ln \beta).
\end{aligned}    
\end{equation}

Having outlined the mapping between simulation outputs and network inputs in log space, we now describe how the training data is generated.
The training process begins by sampling $\ln \beta$, with $\beta$ expressed in units of $1/T$, from a Gaussian distribution with mean 7 and variance 1. This prior choice is only used for this example. When training a network to deal with experimental data, we train a new network using data informed parameter estimates.
For each sampled $\ln \beta$, we compute the corresponding $\beta$ and simulate the dynamics of protein production and division. 
This consists of sampling an initial latent variable, assuming that the cell's last division occurred at a random time drawn uniformly between $0$ and $T$. 
Each simulated cell is then evolved for a duration equivalent to 100 cell division intervals, ensuring convergence to the steady-state protein distribution. 
Between divisions, the number of proteins generated is sampled from a Poisson distribution according to \eqref{constraint1} and at the end they are divided as a Binomial sample as in \eqref{constraint2}.
Further details on how we efficiently parallelize multiple simulations on GPUs are available in our GitHub repository \cite{github}.

The training dataset consists of about a million, $2^{20}$, pairs of $(\ln \beta, \ln G)$ values, divided into 16 batches. 
For each batch, we compute the loss function as defined in \eqref{negloglike} and perform one optimizer step to update the network parameters. 
A full pass over all batches, going once through the entire dataset, defines one epoch. 
The network is trained over many epochs, progressively minimizing the loss and resulting in optimized parameters $\phi^\ast$, as described in Sec.  \ref{sec:training_general}.

To prevent overfitting, at each epoch we regenerate $2^{14}$ new pairs to replace part of the dataset. This practice allows the network to be exposed to different samples in each training iteration, thereby promoting better generalization and reducing the risk of memorizing specific data points. Additionally, we shuffle the dataset indices after every epoch to create varied batch compositions and to further expose the network to a broad range of parameter values during training.

\Specs{
We made the code used to generate the dataset, train the network, and perform all subsequent analyses publicly available on our GitHub repository \cite{github}. Our implementation builds on the \texttt{normflows} package \cite{Stimper23} for normalizing flows. 
Training was performed on a machine equipped with an 8-core Intel Core i7-7700K CPU, 64 GB of system RAM, and an NVIDIA GeForce RTX 3070 GPU with 8 GB of dedicated memory. 
A typical training run, consisting of 1000 epochs, takes approximately 1.5 hours. These hardware specifications apply to all other benchmarks presented in this work.
}

\subsubsection{Inference}\label{sec:model1_results}
\begin{figure}
    \centering
    \includegraphics[width=0.9\linewidth]{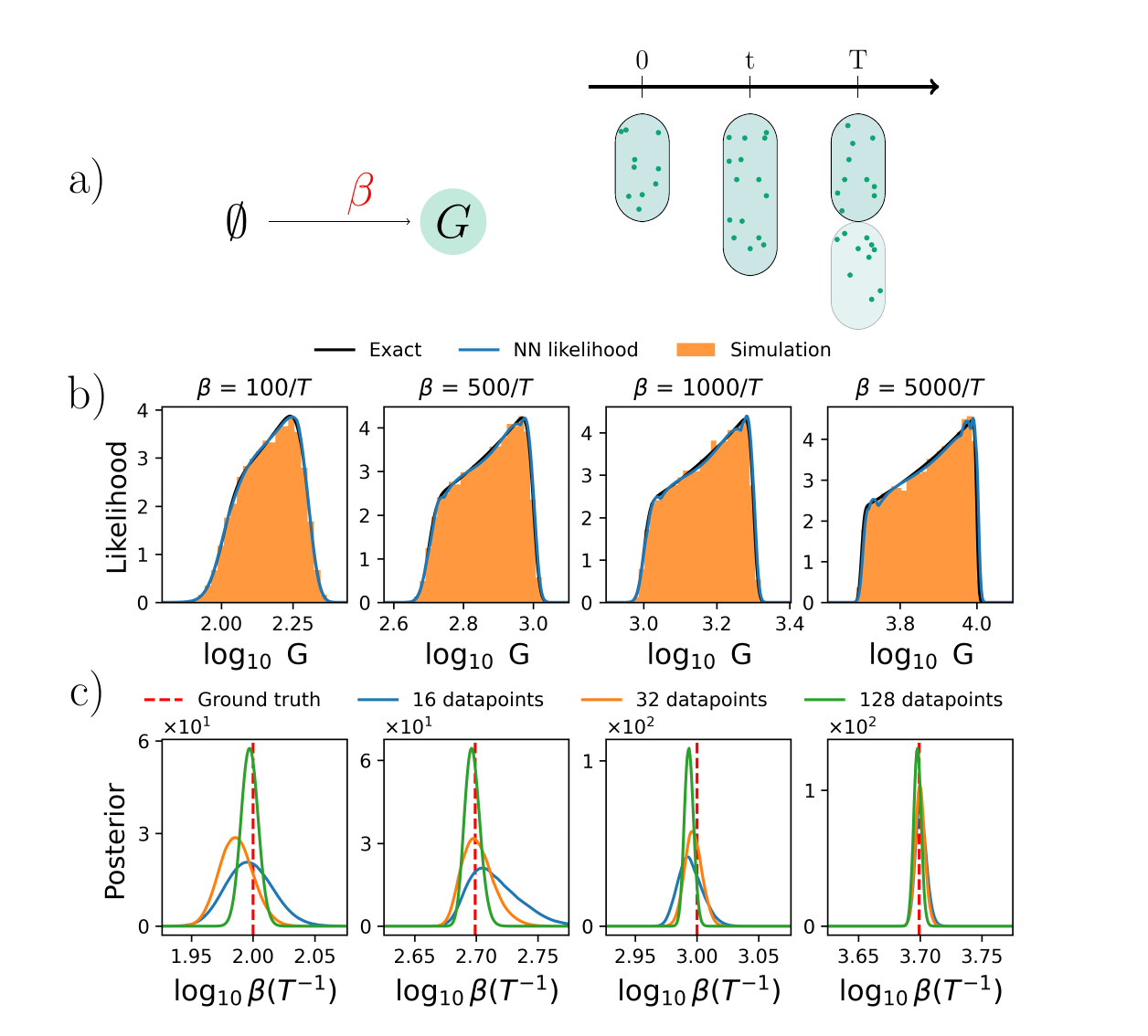}
    \vspace{-.5cm}
    \caption{
    \textbf{Results for Model 1.} 
    a) Schematic of the model described in Sec. \ref{sec:deterministic_cell_division}. The model involves a single parameter, $\beta$, representing the protein production rate. Proteins are diluted during cell division, which occurs at fixed intervals of duration $T$.  
    b) Neural network approximation of the likelihood (blue) for selected $\beta$ values, compared to data simulated using these values as ground truth (orange histogram) and the exact likelihood \eqref{poisson_beta_marginalized} (black). The neural network accurately captures the likelihood across a broad range of $\beta$ values.  
    c) Application of Bayesian inference to estimate $\beta$ from synthetic data. The posterior distribution tightens as the dataset size increases, reflecting improved parameter estimation with more observations. Even with a modest number of observed cells, the inference achieves reasonable accuracy in this example.  
    }\label{fig:deterministic_cell_div_results}
\end{figure}

We present the results obtained using a normalizing flow trained over 2000 epochs in Fig. \ref{fig:deterministic_cell_div_results}.

Fig. \ref{fig:deterministic_cell_div_results}a summarizes the model. In Fig. \ref{fig:deterministic_cell_div_results}b, we compare the neural network's approximation of the likelihood function for selected $\beta$ values to both the exact solution and data simulated using the corresponding $\beta$ values (orange histogram). 
Notably, the neural network was trained on a diverse range of $\beta$ values sampled across the parameter space, rather than specifically on the selected values shown. This comparison demonstrates the network's ability to accurately approximate the likelihood at a given point in the parameter space, even without extensive training data at that specific $\beta$ value.

Finally, in Fig. \ref{fig:deterministic_cell_div_results}c, we illustrate how the neural network–based likelihood approximation can be used for Bayesian inference to estimate $\beta$ from synthetic data. 
We consider sets of $N$ independent snapshots observations $G^n$ collectively denoted simply as$\{G^n\} = \{G^1,G^2,\ldots, G^N\}$, and compute the posterior distribution over a grid of $\beta$ values, with the total likelihood approximated as
\begin{equation}
p(\{G^n\} | \beta) \approx \prod_{n=1}^N \frac{1}{G^n} \, p_{\text{NF}}(\ln G^n | \phi^\ast, \ln \beta),
\end{equation}
where $p_{\text{NF}}$ is the output of the trained normalizing flow model with parameters $\phi^\ast$, obtained from the training scheme described in Sec. \ref{sec:training_model1}, and capture the conditional distribution of $\ln G$ given $\ln \beta$. The posterior is then given by  
\begin{equation}
p(\beta | \{G^n\}) \propto p(\{G^n\} | \beta)\, p(\beta),
\end{equation}
and the proportionality constant is obtained via numerical integration over the grid of $\beta$ values. This posterior reflects how uncertainty about $\beta$ narrows as more data is incorporated. In Fig. \ref{fig:deterministic_cell_div_results}c, we show posterior distributions for increasing dataset sizes $N$, highlighting the gain in precision with larger sample sizes.

\subsection{Model 2  ---  Stochastic cell division}\label{sec:stochastic_cell_division}

In the previous section, we considered a system in which protein production occurs at a constant rate $\beta$, and cell division takes place at regular time intervals $T$. We now extend this model to account for the fact that division timing is inherently stochastic, due to both environmental and intracellular fluctuations. In many practical scenario, such as chemostat experiments, the average division time, $\overline{T}$, can be independently estimated or controlled, as further discussed in Supplemental Information \ref{SIsec:real_data_divisiontime}.
We therefore treat $\overline{T}$ as known and serving as a natural unit of time.   

While a comprehensive treatment of mechanisms causing variability in cell division timing is beyond the scope of the present work\cite{Biswas18,LeTreut2021,Sauls16}, we adopt a phenomenological approach in which division times are modeled as the sum of a large number of sequential, memoryless steps \cite{Golubev16,Chao19,Belluccini22}. 
This framework leads naturally to the use of Gamma distributions. 

Thus, we model division times as Gamma distributed in units of the average division time, $\overline{T}$, that is
\begin{equation} \label{time_dist} 
\frac{T}{\overline{T}} \sim \text{Gamma}\left(\frac{1}{\sigma^2}, \sigma^2\right),
\end{equation} 
where we introduced a new parameter, $\sigma$, equivalent to the standard deviation of $T/\overline{T}$.
This formulation allows us to flexibly capture biologically realistic fluctuations in division timing while maintaining analytical control over the statistical properties of the system.
 
\begin{figure}
    \centering
    \includegraphics[width=0.9\linewidth]{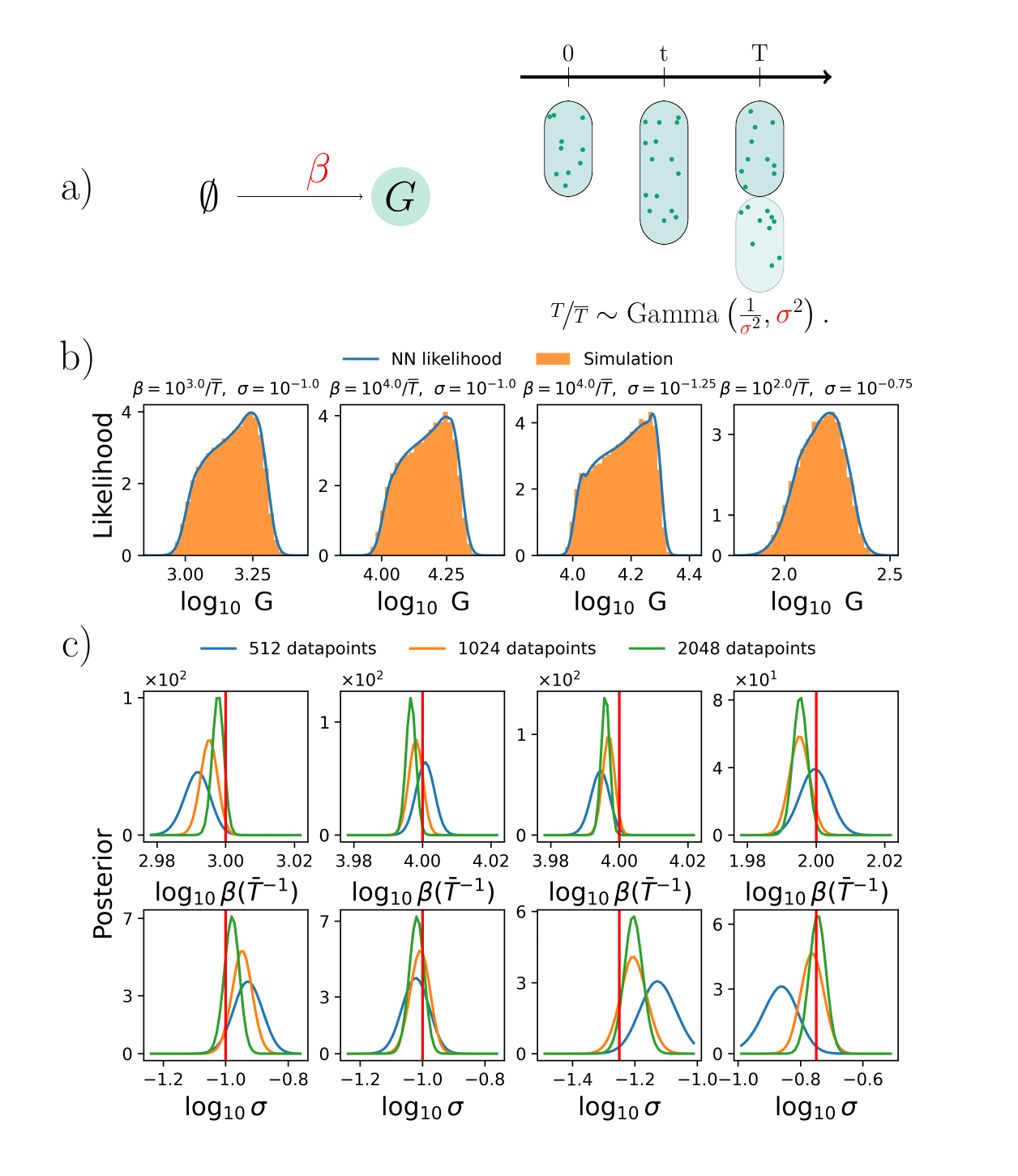}
    \vspace{-1cm}
    \caption{
    \textbf{Results for Model 2.} 
    a) Schematic of the model described in Sec. \ref{sec:stochastic_cell_division}. We learn the pair of parameters $\theta = \{\beta,\sigma\}$  representing the protein production rate and cell division time standard deviation. 
    b) Neural network approximation of the likelihood (blue) for selected $\theta$ values, compared to data simulated using these values as ground truth (orange histogram). The neural network accurately captures the likelihood across a broad range of $\beta$ values.  
    c) Application of Bayesian inference to estimate both parameters from synthetic data. The posterior distribution tightens as the dataset size increases, from 512 to 2048 cells, reflecting improved parameter estimation with more observations. 
    }\label{fig:stochastic_cell_div_results}
\end{figure}

In this framework, the parameters to be inferred include the protein production rate, $\beta$, and the standard deviation of the cell division time, $\sigma$, which captures the stochasticity of the cell cycle as defined in \eqref{time_dist}. 
As in Model 1, the number of proteins produced during a time interval $\Delta t$ between cell divisions is Poisson distributed with mean $\beta \Delta t$ and upon each division event the number of proteins in the cell is reduced according to a Binomial  distribution with success probability $1/2$. A summary of the model is presented in Fig. \ref{fig:stochastic_cell_div_results}a.

For this model, deriving an exact analytical expression for the steady-state protein distribution, analogous to  \eqref{poisson_beta_marginalized}, becomes infeasible. Consequently, we rely on simulation-based methods combined with normalizing flow-based likelihood approximation to perform parameter inference, following the framework introduced in the previous section.

\subsubsection{Training the neural network}

The training process follows a structure similar to that used in the deterministic case of Model 1 (Sec. \ref{sec:training_model1}), but now incorporates stochastic division times. 
Specifically, the neural network is trained to approximate the probability distribution $p(G | \beta, \sigma)$. During training, each parameter set, $\theta = (\ln \beta, \ln \sigma)$, is sampled, where $\ln \sigma$ is drawn from a Gaussian distribution with mean $-2.3$ and standard deviation $0.5$, and $\ln \beta$ is drawn from a Gaussian with mean $7$ and standard deviation $1$ (these priors are used only for training on synthetic data). 
As before, the inputs to the network are $\ln G$ and $\theta$, and the output is the learned likelihood function $p_\text{NF}(\ln G | \phi^\ast,\theta)$. 

As in the deterministic case in Model 1, we again generate a dataset of $2^{20}$ pairs of $(\theta, \ln G)$, split into 16 batches. The network is trained iteratively using the negative log-likelihood loss function, as defined in \eqref{negloglike}. At each epoch, $2^{14}$ new pairs are sampled to replace a fraction of the dataset, and the data indices are shuffled to ensure diverse batch compositions.
In the same machine specifications as described in Sec. \ref{sec:model1_results}, the \Specs{training consisting of  1000 epochs using the code available on GitHub took approximately 2 hours.}

\subsubsection{Inference}

We now evaluate the performance of the trained normalizing flow model in approximating the likelihood function in {Model 2, modeling stochastic division dynamics}. The results are summarized in Fig. \ref{fig:stochastic_cell_div_results}.

Fig. \ref{fig:stochastic_cell_div_results}a provides an overview of the model described in Sec. \ref{sec:stochastic_cell_division}, highlighting the two parameters to be learned ($\beta$ and $\sigma$).
In Fig. \ref{fig:stochastic_cell_div_results}b, we compare the neural network's likelihood approximation and simulations at these parameter values. 
Finally, in Fig. \ref{fig:stochastic_cell_div_results}c, we apply Bayesian inference to estimate both $\beta$ and $\sigma$ from synthetic data. The posterior distributions illustrate how the inferred values become more constrained as more observations are incorporated. This result highlights the effectiveness of the neural network-based likelihood approximation in supporting inference for stochastic division models.  

\subsection{Model 3  ---  Stochastic cell division with two-state regulation and uneven partitioning}\label{sec:fc}

We now extend our model by introducing additional layers of biological complexity.
With the intent of applying this method to flow cytometry data of \emph{S. cerevisiae}, we incorporate a two-state regulatory mechanism governing protein production, uneven cell division (as observed in yeast), and the indirect measurement of protein content via flow cytometry.
Due to the difficulties introduced by these additional biological features, Model 3 does not admit an analytical solution for $p(G|\theta)$.

In this extended framework, cells transition between two internal states: an active gene state, where protein production occurs at rate $\beta$, and an inactive state, where the protein of interest, whose counts are observed indirectly through fluorescence, is not produced. 
Simulating this model requires keeping track of both the number of proteins and the latent internal gene state over time.
The time until the next gene state switch is exponentially distributed, with transition rates given by the activation rate $\lambda_{\text{act}}$ (inactive to active) and the inactivation rate $\lambda_{\text{ina}}$ (active to inactive).
The parameters to be inferred now include $(\beta, \sigma, \lambda_{\text{act}}, \lambda_{\text{ina}})$.

As with the previous model, cell division times $T$ are sampled from \eqref{time_dist}, while the times to state switching events are sampled from Exponential distributions determined by the corresponding transition rates.
We assume that the  production of the protein of interest occurs only during the active gene state.
Thus, within any time interval $\Delta t$, defined as the time between cell division or state switching events, the number of proteins of interest produced is modeled as a Poisson random variable with mean $\beta \Delta t$.
 
Unlike in {Model 2}, we do not assume that cells divide symmetrically. Rather, we model the volume fraction $\rho$ of the daughter cell relative to the total volume as a Beta-distributed random variable. Drawing on previous chemostat fermentation data at various dilution rates reported by some of us \cite{Martinez22}, we have $\rho \sim \text{Beta}(6, 14)$, corresponding to a mean of 0.3 and a standard deviation of 0.1.
During each division event in the simulation, the cell is randomly assigned to be either a mother or a daughter and the number of proteins inherited is determined by a Binomial distribution with success probability $\rho$ (if the cell is assigned as the daughter) or $1-\rho$ (if assigned as the mother).
In other words, proteins are partitioned independently, with inheritance probabilities proportional to the corresponding cell’s volume fraction.

Although previous models assumed direct  count measurements for the protein of interest, here we model protein abundance as being correlated with fluorescence intensity. In this framework, we must account for autofluorescence: a background signal emitted by cells at the same frequency as the protein marker.

The autofluorescence intensity per cell, denoted by $I_\text{af}$, is approximated using a fixed normalizing flow trained on single-cell fluorescence measurements collected under minimal stress conditions (see Supplemental Information \ref{SIsec:autofluo}). This allows us to model the observed fluorescence $I$ as the sum of the protein fluorescence signal $I_g$ and the autofluorescence contribution: \begin{equation}
\label{Isum} I = I_\text{af} + I_g. 
\end{equation}

To incorporate fluorescence measurements into our model, we first sample $I_\text{af}$ from the pre-trained normalizing flow. Next, the fluorescence signal $I_g$ {of the protein of interest} is generated according to: \begin{equation}
\label{Ig_poisson} I_g \sim \text{Gaussian}(\xi G,\xi G), 
\end{equation} 
where $\xi$ is a calibration factor reflecting the  conversion of fluorescent particles ($G$) into total fluorescence intensity ($I_g$). $\xi$ was determined from bead experiments (see Supplemental Information \ref{SIsec:intensityfluo}).

Finally, the observed fluorescence $I$ is obtained as \eqref{Isum}, and the neural network takes as input the log-transformed intensity $\ln I$ along with the logarithm of the parameters $\psi_\theta$. The output is the learned likelihood function $p_\text{NF}(\ln I \mid \phi^\ast,\theta)$.

\subsubsection{Training the neural network}

The training process follows a structure similar to the previous examples in Sec. \ref{sec:deterministic_cell_division} and \ref{sec:stochastic_cell_division}. Each parameter set is sampled as 
 $\theta = (\ln \beta, \ln \sigma, \ln \lambda_{\text{act}}, \ln \lambda_{\text{ina}})$, 
where  $\ln \sigma $ is drawn from a Gaussian distribution with mean  $-2.3$ and standard deviation  $0.5$, 
$\ln \beta $ is sampled from a Gaussian distribution with mean  $7$ and standard deviation $1$, $\ln \lambda_\text{act}$ with mean $1$ and standard deviation $1.5$, and $\ln \lambda_\text{ina}$ with mean $-1$ and standard deviation $1.5$.
These priors apply only to synthetic data; later when applying to real data, we retrain the network using calibrated estimates as detailed in Sec. \ref{sec:real}.

To ensure robust training, we again generate a dataset of  $2^{20} $ pairs of  $(\ln I, \theta) $ split into 16 batches. The network is trained iteratively using the {negative log-likelihood loss function} defined in  \eqref{negloglike}. At each epoch,  $2^{14} $ new pairs are sampled to replace a fraction of the dataset. \Specs{Training the model for 1000 epochs using the code available on GitHub took approximately 2.5 hours.}

\subsubsection{Inference}\label{sec:fc_results}
We now evaluate the performance of the normalizing flow model in approximating likelihoods and enabling Bayesian inference in this more complex setting, which includes a two-state gene activation regulatory mechanism, uneven protein partitioning, and fluorescence-based measurements.

\begin{figure}
    \centering
    \includegraphics[width=0.9\linewidth]{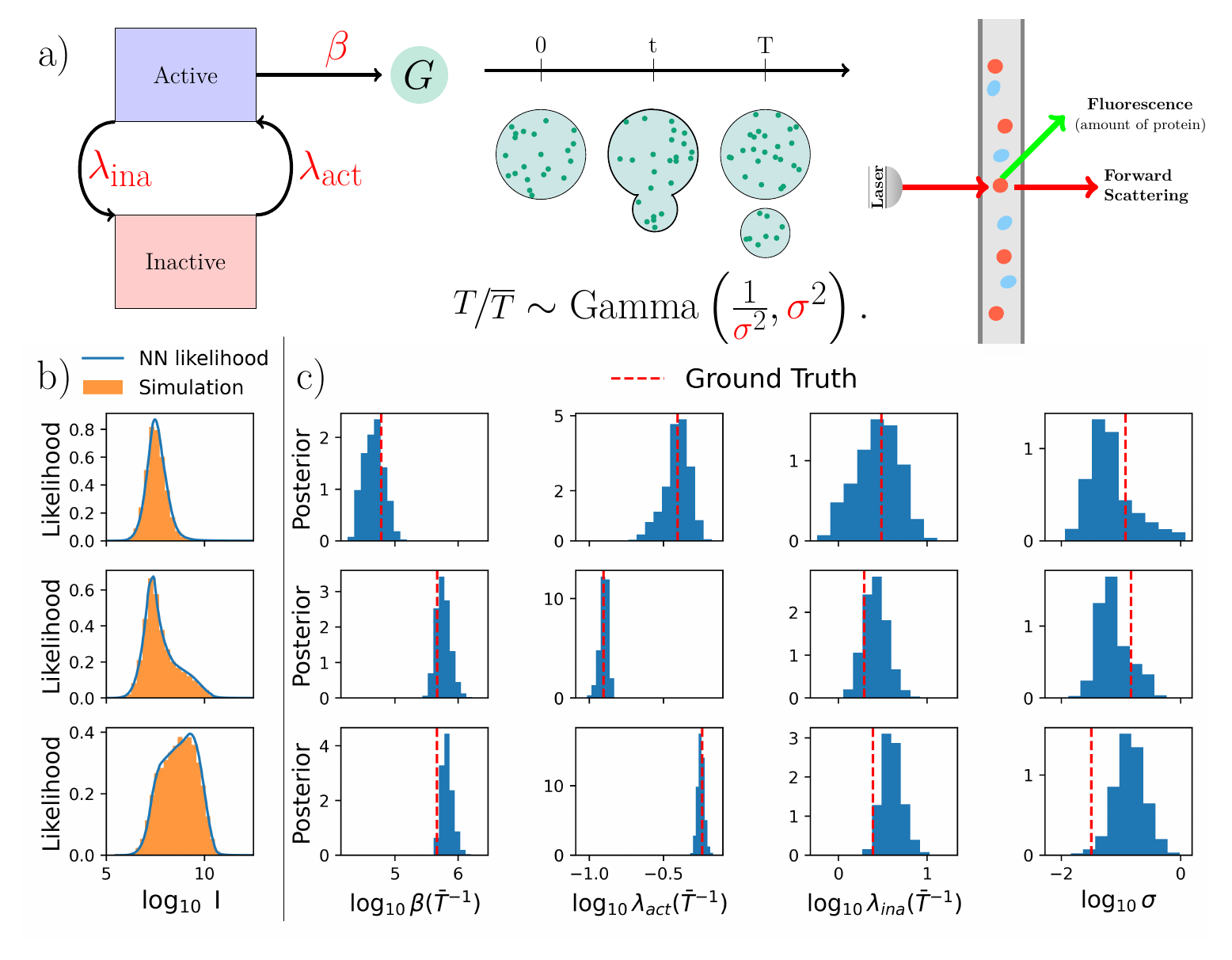}
    \vspace{-1cm}
    \caption{
    \textbf{Results for Model 3.} 
    a) Schematic of the model described in Sec. \ref{sec:fc}. Including the budding-type division that is common in \emph{S. cerevisiae}. We learn the set of parameters $\theta = \{\beta, \sigma, \lambda_{\text{act}}, \lambda_{\text{ina}}\}$ representing the protein production rate, cell division time standard deviation, activation rate, and deactivation rate. 
    b) Neural network approximation of the likelihood function for the total flow cytometry intensity $I$, in arbitrary florescence units, shown in blue for selected parameter values $\theta$. The orange histograms represent samples drawn from simulations of the model using these parameter values.  
    c) Bayesian inference results for parameter estimation. Each column corresponds to a different parameter, with inferred posterior distributions shown as blue histograms obtained via MCMC. The red dashed lines indicate the ground truth values used to generate the synthetic data. These posteriors are obtained with 2048 datapoints. Each row in c) corresponds to the same ground truth values used in the respective panel in b). 
    }\label{fig:fc_results}
\end{figure}

Fig. \ref{fig:fc_results}a provides an overview of this extended model. Unlike previous sections, where cell division was modeled as a symmetric split (bacterial division) this model accounts for the asymmetric division of yeast, which occurs through budding. In this process, daughter cells inherit different amounts of protein depending on their size at division.

The increased number of parameters $(\beta, \sigma, \lambda_{\text{act}}, \lambda_{\text{ina}})$ makes direct posterior evaluation across a range of values in the region of interest computationally impractical. While this approach was feasible in the example of Sec. \ref{sec:deterministic_cell_division}, which involved a single parameter $\beta$, and in Sec. \ref{sec:stochastic_cell_division}, which considered two parameters $\beta$ and $\sigma$, it becomes intractable in higher dimensions. We therefore employ a posterior sampling though an Markov chain Monte Carlo MCMC method with an adaptive proposal strategy inspired by \cite{Haario00}, which dynamically refines the sampling distribution for efficiency.
At each MCMC step, we evaluate the posterior by substituting the true likelihood with its approximation learned via a normalizing flow, as described in Sec. \ref{sec:training_model1}.
Further details on this implementation are provided in Supplemental Information \ref{SIsec:MCMC}.

In Fig. \ref{fig:fc_results}b, the learned likelihood functions exhibit qualitatively distinct patterns depending on the parameter values. Despite this variability, a single neural network successfully approximates the likelihood across a broad range of conditions, demonstrating its flexibility and accuracy.

Furthermore, Fig. \ref{fig:fc_results}c shows that the inferred posterior distributions are well-separated for different ground truth values, confirming the neural network's ability to resolve distinct kinetic regimes. This separation highlights the effectiveness of the method in distinguishing parameter variations, even within a complex biological system.

With this trained network incorporating realistic biological features relevant to \emph{S. cerevisiae}, we now apply it to experimental flow cytometry data to infer in vivo protein kinetics.

\subsection{Learning kinetics from flow cytometry of \emph{S. cerevisiae}}
\label{sec:real}

As a real-world application of the model developed in the previous sections, we analyze flow cytometry data from a study previously conducted by some of us \cite{Martinez22}, which examined the general stress response of \emph{S. cerevisiae} under continuous cultivation. This yeast species is known to employ bet-hedging strategies to balance growth and stress resistance \cite{Levy12}. As a result, even under optimal growth conditions, genetically identical cells may exhibit divergent phenotypes with some continuing to grow while others invest in stress resistance, thus leading to significant heterogeneity in gene expression.

To monitor this heterogeneity, a transcriptional reporter was developed using GFP under the control of the promoter of the \emph{glc3} gene, which is involved in glycogen synthesis and is known to be modulated by nutrient limitation \cite{Martinez22,Zid14}. This reporter was previously shown to be a marker of bet-hedging in yeast \cite{Levy12}.

To systematically vary nutrient availability, chemostat experiments were conducted at three dilution rates: 0.12, 0.23, and 0.33$h^{-1}$. At the lowest rate, nutrient limitation is high and the \emph{glc3} reporter is strongly expressed. At the highest rate, nutrients are abundant, suppressing reporter expression. The intermediate condition is expected to yield a heterogeneous mix of stressed and growing cells. During these cultivations, flow cytometry was used to quantify GFP expression in individual cells \cite{Martinez22,Henrion23}.
 
Here flow cytometry analyzes individual events, where each event corresponds to a single particle, usually a single cell, passing through a laser beam. For this analysis, we focus on forward scatter (correlating with cell size \cite{Konokhova13}), side scatter (providing orthogonal size information), and FL1 fluorescence intensity quantifying the amount of GFP, a proxy for the expression of the \emph{glc3}.

Flow cytometry data, however, can be impacted by artifacts. Debris and non-cellular particles may be misclassified as valid events (cells).  
To ensure data quality, we apply a preprocessing pipeline  removing those artifacts.  
Following recomendation from Flow cytometer manuals \cite{BDAccuri2019}, we first impose a lower cutoff on forward scatter is used to exclude non-cellular particles. Then, a support vector machine classifier \cite{svm_scikit_learn} is employed to detect and remove outliers, as validated in other studies of both flow and mass cytometry \cite{Saeys16,Kimball18}. This procedure discards only a small fraction of events but significantly improves the reliability of downstream inference. Full details are provided in Supplemental Information \ref{SIsec:SVM}.

The resulting fluorescence intensity distributions are shown in Fig. \ref{fig:fc_results}a. These distributions support the assumption, further discussed in Supplemental Information \ref{SIsec:Real_data_active} that only one gene state is primarily responsible for protein production.
 
In the chemostat, the dilution rate determines the rate of nutrient resupply and thus serves as a proxy for cellular stress. Lower dilution rates increase nutrient limitation, activating stress responses. Among the tested conditions, 0.33 $h^{-1}$ is the highest dilution rate at which a stable population could be maintained. As nutrient levels are high and stress is minimal, this condition is used to measure autofluorescence, as detailed in Supplemental Information \ref{SIsec:autofluo}. The dilution rate of 0.23$h^{-1}$ represents low stress, while 0.12$h^{-1}$ corresponds to high stress and prolonged nutrient limitation.
 
Since the cell division time in a chemostat is the inverse of the dilution rate, see Supplemental Information \ref{SIsec:real_data_divisiontime}, we analyze data in rescaled time units such that the average cell cycle division time is the unit of time. 
Inference is performed in these units and later converted back to hours, yielding average division times of approximately 4.35$h$ (low stress) and 8.33$h$ (high stress).

\begin{figure}[t]
    \centering
    \includegraphics[width=0.9\linewidth]{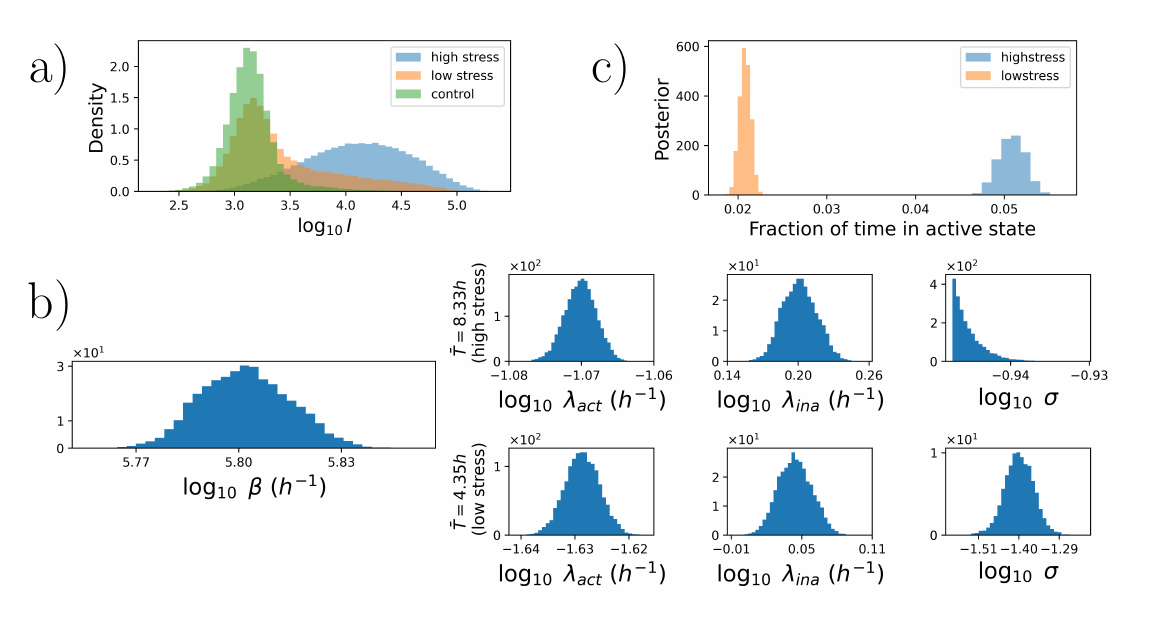}\vspace{-.6cm}
    \caption{
    \textbf{Inference of protein kinetics from real flow cytometry data.}
    (a) Fluorescence intensity, in arbitrary florescence units, distributions for yeast populations at different dilution rates. Since lower dilution rates correspond to higher nutrient limitation and thus higher stress, the high stress condition (0.12 $h^{-1}$) shows significantly elevated fluorescence compared to the low stress condition (0.23 $h^{-1}$), while the control (0.33 $h^{-1}$), used to calibrate autofluorescence, exhibits only background signal.
    (b) Posterior distributions of the inferred kinetic parameters obtained through MCMC. The model assumes a constant protein production rate ($\beta$), while gene activation ($\lambda_{\text{act}}$), inactivation ($\lambda_{\text{ina}}$), and relative division-time variability ($\sigma$) are different at each stress condition.
    (c) Estimated fraction of time spent in the active state, computed via \ref{fraction}. Cells in high stress are active $\approx 5\%$ of the time, versus $\approx 2\%$ in low stress. Although activation is brief, proteins persist across several cell cycles, explaining the high observed fluorescence.
    }
    \label{fig:real_results}
\end{figure}

As shown in Fig. \ref{fig:real_results}a, fluorescence intensity is orders of magnitude higher in the high stress condition compared to low stress and the autofluorescence control. At first glance, this may suggest that cells under high stress remain constantly in the active state, producing fluorescent proteins continuously. However, a deeper analysis of the inferred kinetic parameters reveals a more nuanced picture.

In the network training performed on synthetic data, we sampled model parameters $\theta$ from a predefined distribution. 
However, without prior knowledge of which parameter values are actually compatible with the real data, this choice may lead to suboptimal training. 
Ideally, we want to train the neural network around parameter regions where the likelihood is expected to be high. This approach ensures that the network sees more examples of datasets in regions similar to the real data, thereby leading to better approximations of the likelihood function. 
At the same time, it is important to maintain sufficient flexibility, meaning including a wide enough range of parameter values to capture possible biological variability and account for uncertainties. 
To achieve this balance in way compatible with the data in Fig. \ref{fig:real_results}a, we train the neural network on data generated from parameters calibrated to the real experiment using a more naive form of SBI, as detailed in Supplemental Information  \ref{SIsec:precalibration_real}.

We then perform MCMC inference, similarly to how it was done in Sec. \ref{sec:fc_results} (with further details in Supplemental Information \ref{SIsec:MCMC}), and the posterior obtained is presented in Fig. \ref{fig:real_results}b. The estimated protein production rate, $\beta$, is assumed unknown and to be learned but is assumed to be the same across both stress conditions. 
This is because $\beta$ is set by the genetic machinery (\emph{e.g.}, mainly related to the ribosome binding) and is not directly tied  to environmental factors, while the other parameters: cell division variability ($\sigma$), gene activation rate ($\lambda_{\text{act}}$), and inactivation rate ($\lambda_{\text{ina}}$) are condition dependent and differ across high and low stress conditions.

From these inferred parameters, we compute the fraction of time spent in the active state, given by
\begin{equation}\label{fraction}
    f_\text{active} = \frac{\lambda_\text{act}}{\lambda_\text{act}+\lambda_\text{ina}} .
\end{equation}
The resulting estimates, shown in Fig. \ref{fig:real_results}c, reveal that cells under high stress actually spend only about 5\% of the time in the active state, compared to 2\% in the low stress condition. This finding suggests that under high stress, cells are not persistently active but instead switch into the active state only briefly. However, due to the slow degradation of the produced proteins, fluorescence levels remain elevated across multiple cell cycles, leading to the observed high fluorescence intensity in Fig. \ref{fig:real_results}a.  
This underscores the importance of modeling cell division explicitly and the necessity of the current framework: a na\"{i}ve interpretation of raw fluorescence would incorrectly suggest that a large fraction of cells were active. In contrast, capturing cell division as part of the inference reveals that gene activation is rare and short-lived. Our framework thereby helps us disentangle inherited fluorescence from true expression dynamics.

\section{Conclusion}

We considered the challenge of estimating protein production, even extending to the case of gene switching kinetics, when cells are dividing. In this case, the instantaneous protein count within each cell depends on its entire cell-division history. 

We applied this framework to real flow cytometry data from \emph{S. cerevisiae} cultured in chemostats at different dilution and thus nutrient concentration rates (Fig. \ref{fig:real_results}). The results show that contrary to expectation --- and although fluorescence intensity is significantly higher under high-stress conditions --- cells rarely find themselves in a \emph{glc3} upregulated state \cite{Martinez22,Henrion23} leveraging instead gene product inheritance through parent cells. Interpreting these results while ignoring the cell cycle would lead to fundamentally different conclusions: \emph{glc3} is frequently upregulated ignoring altogether gene products inherited through division.

Indeed, our treatment now extends approaches that so far have made the simplifying assumption of modeling cell division as an additional chemical reaction 
\cite{Friedman06, Shahrezaei08, Pendar13, Kumar14, Chen22, Bingjie24, Jia17, Jiao2024} implicitly assuming that reduction in protein numbers arises solely from individual degradation rather than partitioning.
As shown in Fig. \ref{fig:simulated_forward}, such approximations can lead to qualitatively incorrect likelihood distributions. In contrast, our method models cell division times explicitly and performs likelihood-based inference directly from forward simulations, thereby avoiding these oversimplifications.

Learning protein production and gene switching kinetics requires we move beyond existing physics-inspired inference schemes relying on likelihoods which, in turn, often reduces to solving the coinciding master equation \cite{Pessoa24, SzavitsNossan24, Kilic23, Munsky06}, Fokker-Planck \cite{Bryan20,Reich21,Tabandeh22} and other often equations which often cannot be solved or  easily formulated though the process may be simulated. 

In this work, we motivated how neural networks can be leveraged as approximators to likelihoods using data from forward simulations for both common (yet intractable non-Markovian) problems but, potentially, also for problems where the likelihood is computationally expensive without necessarily being fundamentally intractable. 
In such settings, the method presented here provides a practical surrogate preserving the statistical framework of likelihood-based inference while substantially reducing computational costs.

While we have opted for normalizing flows in this work due to their straightforward probabilistic formulation, other neural network architectures whose outputs are tractable probability distributions (such as score-based diffusion models \cite{Song21}) may also have been used. For example, diffusion models have been applied to infer Boltzmann weights for proteins \cite{pepflow}, offering a promising alternative generative modeling framework.

\section*{Acknowledgments}
All authors thank Roberto Covino for inspiration in simulation-based inference and encouraging us to think deeply about this new approach.
SP acknowledges support from the NIH (R35GM148237), ARO (W911NF-23-1-0304), and NSF (Grant No. 2310610).

\bibliographystyle{elsarticle-num} 
\bibliography{refs}

\newpage
\appendix
\renewcommand{\thesection}{S\arabic{section}}
\renewcommand{\thesubsection}{S\arabic{section}.\arabic{subsection}}
\renewcommand{\theequation}{S\arabic{equation}}
\renewcommand{\thefigure}{S\arabic{figure}}
\renewcommand{\thetable}{S\arabic{table}}
\renewcommand{\thepage}{S  \arabic{page}}

\setcounter{section}{0}
\setcounter{equation}{0}
\setcounter{figure}{0}
\setcounter{table}{0}
\setcounter{page}{1}

\section*{Supplemental Information}

\section{Why we can assume we know the average cell division time}\label{SIsec:real_data_divisiontime}
This Supplementary Information section explains why we assume the mean cell division time $\overline{T}$ is known in Sec. \ref{sec:stochastic_cell_division} and Sec. \ref{sec:fc}.

In the chemostat experiments for \emph{S. cerevisiae}, as shown in Sec. \ref{sec:real}, the average division rate fueled by continuously supplied nutrients matches the {dilution rate} $D$ of the chemostat. 
This is achieved through adaptation. Briefly, adaptation occurs by virtue of constraints on growth:  if cells divide too slowly, the population is washed out; if they divide too quickly, nutrient depletion or crowding will suppress further growth. This balance can be captured by a simple ordinary differential equation describing the population size $N(t)$:
\begin{equation}
\frac{dN}{dt} = \mu N - D N,
\end{equation}
where $\mu$ is the specific growth (division) rate of the cells. {At steady state, $\frac{dN}{dt} = 0$, we must have $\mu = D$.} Since the mean division time, $\overline{T}$, is the inverse of the division rate, it follows that $\overline{T} = \nfrac{1}{D}$. 

{For this reason, throughout our work, we have adopted units where the mean division time $\overline{T}$ defines the unit of time, setting $\overline{T} = 1$. All results are thus expressed in terms of $\overline{T}$. In Sec. \ref{sec:real}, when analyzing real experimental data, we appropriately convert these dimensionless results back into physical units (hours) by dividing all rates (which have units of inverse time) by $D$.}

\newpage
\section{Autofluorescence calibration}\label{SIsec:autofluo}

The distribution of autofluorescence intensity, $I_\text{af}$, introduced in Sec. \ref{sec:fc}, appears in Eq. \eqref{Isum}, but its sampling procedure was not previously specified.

Among the tested conditions, 0.33 $h^{-1}$ represents the highest dilution rate at which a stable population could be maintained. Since the fluorescent reporter is driven by a stress-response gene, this condition—with continuous nutrient availability and minimal stress—is assumed to reflect pure autofluorescence. It serves as the reference for autofluorescence calibration, as detailed in Supplemental Information \ref{SIsec:autofluo}.

The distribution for $I_\text{af}$ is modeled using a normalizing flow trained to approximate the distribution of intensities observed at 0.33 $h^{-1}$. This normalizing flow uses the same architecture as those presented in the main text but is trained unconditionally—that is, it learns a probability distribution without any conditioning variables. Further implementation details are available on GitHub \cite{github}.

The data from the 0.33 $h^{-1}$ dilution condition, like the others, was preprocessed as described in SI \ref{SIsec:SVM}.

\begin{figure}[h]
    \centering
    \includegraphics[width=0.6\textwidth]{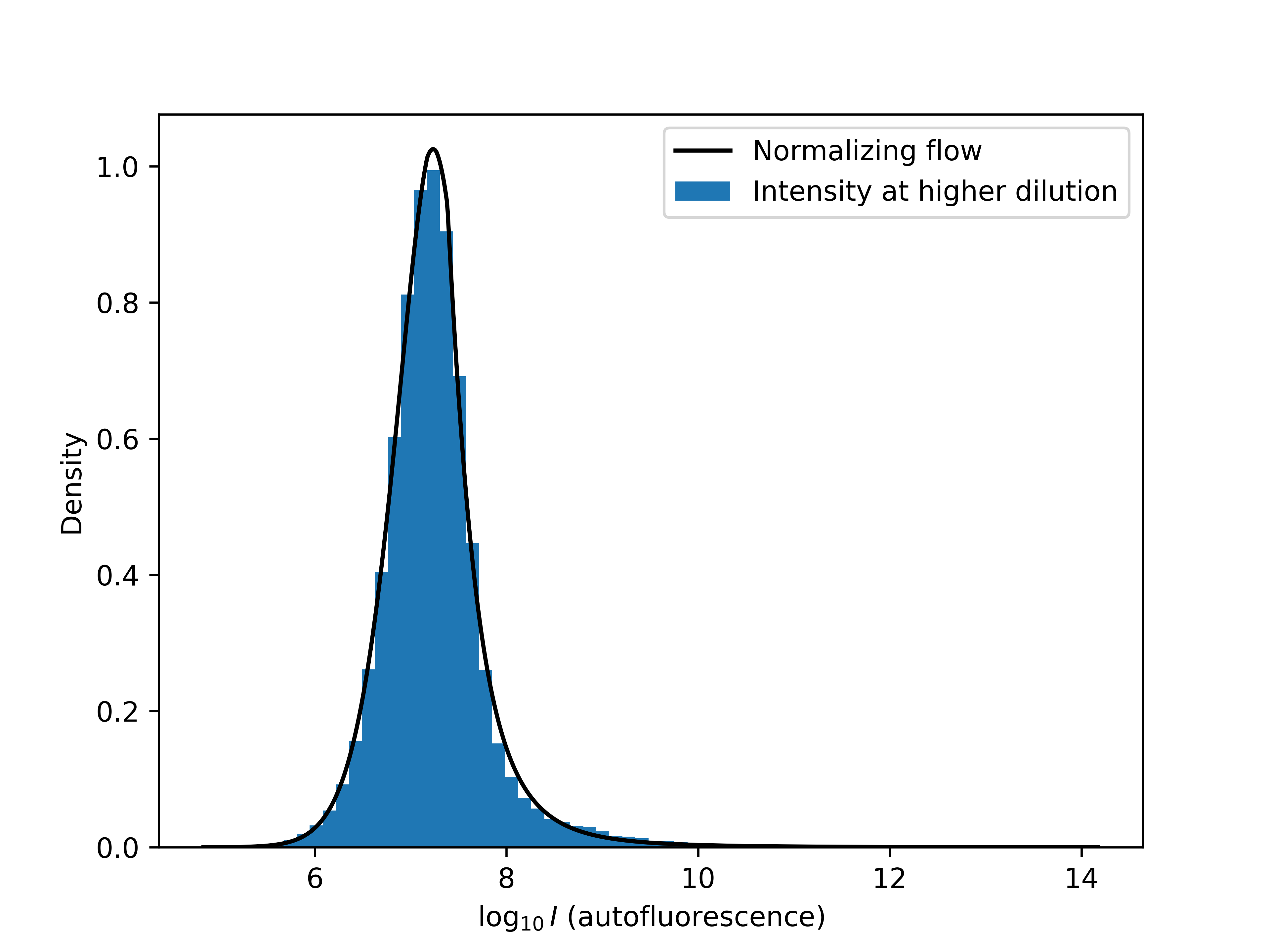}
    \caption{
        Histogram of fluorescence intensities at dilution rate 0.33 $h^{-1}$, representing a low-stress control condition dominated by autofluorescence, equivalent to the ``control" histogram in Fig. \ref{fig:real_results}a. 
        This distribution was used to train the normalizing flow model defining the autofluorescence intensity distribution $I_\text{af}$ used in \eqref{Isum}.
    }
    \label{autofluo}
\end{figure}

\newpage
\section{Calibration of fluorescence intensity}
\label{SIsec:intensityfluo}

{To a attest the relationship of the measured florescence in the FC and the number of fluorescent proteins in the cell we used a set of AcGFP1/EGFP flow cytometer calibration beads (Takara Bio, San Jose,CA,cat:632594). This set consists of a mixture of six discrete bead populations having different fluorescent intensities. Each bead population fluorescent intensity is calibrated to a specific molecular equivalent of soluble fluorofor (MESF). Therefore this beads populations were used as the stardards for our intended calibration.}

The observed fluorescence intensity $I_g$ is assumed to be proportional to the number of fluorescent proteins $G$ in the cell. As described in the main text, we model this relationship as
\begin{equation}\label{Ig_poisson}
    I_g \sim \text{Normal}(\xi G,\xi G ),
\end{equation}
where $ \xi $ is a calibration factor that relates molecular counts to fluorescence units. This factor was experimentally determined using calibration beads with known fluorophore content.

Specifically, we used a commercial bead mixture containing six distinct bead populations, each with a known equivalent number of GFP molecules, as provided by the manufacturer. The bead sample was measured under the same flow cytometry settings as the experimental data. We then fit a mixture of six Gaussian distributions to the measured intensity histogram, corresponding to the six known bead populations.

From this fit, we extracted the mean fluorescence intensity of each population. A linear regression was then performed between these mean intensity values and the manufacturer-specified equivalent number of fluorophores. The resulting slope gives the calibration factor $ \xi $, which we estimate to be
$\xi = 0.0533 \quad \text{fluorescence units per fluorophore}.$

\newpage
\section{MCMC strategy}\label{SIsec:MCMC}
\paragraph{MCMC procedure}
In this section, we describe the Markov Chain Monte Carlo (MCMC) procedure referenced in Secs. \ref{sec:fc_results} and \ref{sec:real}. Although the two applications differ in terms of model and data (with the results in Sec. \ref{sec:real} corresponding to the real data setting), both implementations follow a common adaptive MCMC framework. We first present the shared structure of the algorithm, followed by additional details specific to each case.

In both cases, the posterior distribution $p(\theta | \text{data})$ is approximated using a normalizing flow, as described in the main text. The flow allows for efficient and differentiable evaluation of the likelihood, which is essential for the computation of the log-posterior during sampling.

\paragraph{Initialization.} Prior to sampling, we performed a preliminary search to identify a suitable starting point. We sampled 1,000 parameter vectors $ \theta^{(i)} $ from the prior distribution and evaluated their unnormalized log-posterior. Each sample was then progressively rescaled toward the current best parameter $ \theta^* $ using
\begin{equation}
\theta_{\text{prop}} = \theta^* + \frac{1}{\alpha}(\theta^{(i)} - \theta^*), \quad \alpha = 1, \dots, 10.
\end{equation}
If a proposed $ \theta_{\text{prop}} $ yielded a higher posterior value, it replaced $ \theta^* $. This procedure concentrated the search near high-probability regions and produced an initial point close to the posterior mode.

\paragraph{Adaptive phase.} We then applied a Metropolis-Hastings MCMC algorithm to generate samples from the posterior distribution $ p(\theta \mid \text{data}) $. At each step, a candidate parameter $ \theta_{\text{prop}} \sim q(\cdot \mid \theta) $ was drawn from a symmetric Gaussian proposal centered at the current value. The proposed value was accepted with probability
\begin{equation}
A(\theta_\text{prop},\theta) = \min\left(1, \frac{p(\theta_{\text{prop}} | \text{data})}{p(\theta | \text{data})} \right),
\end{equation}
and otherwise the chain remained at $ \theta $.

Inpired by Haario \cite{Haario00}, to improve sampling efficiency, the proposal covariance matrix $ \Sigma $ was adapted during the early phase of sampling. Initially, $ \Sigma $ was set to a small isotropic matrix. Every 150 iterations, $ \Sigma $ was updated using the empirical covariance of the most recent 200 samples:
\begin{equation}
\Sigma = \frac{2.4^2}{d} \left( \mathrm{Cov}[\theta^{(t)}] + \epsilon I \right),
\end{equation}
where $ d $ is the dimensionality of $ \theta $, and $ \epsilon = 10^{-8} $ is a regularization term to ensure numerical stability. Adaptation was performed only if the acceptance rate fell outside the target range of 20--50\%. Once the acceptance rate remained within this range for 10 consecutive adaptation steps -- meaning that $ \Sigma $ was no longer being updated -- the proposal distribution was fixed for the remainder of the chain. This adaptive strategy allowed the proposal to better reflect the geometry of the posterior, improving convergence while maintaining detailed balance.

\paragraph{Sampling stage.} After adaptation was complete, we continued sampling using the final proposal distribution. From this stabilized phase of the chain, we collected 10,000 samples to approximate the posterior distribution over parameters. Only these post-adaptation samples are shown in the results presented in Figures~\ref{fig:fc_results} and~\ref{fig:real_results}.

\paragraph{MCMC initialization in Sec. ~\ref{sec:real}.} For the MCMC results presented in Sec. ~\ref{sec:real}, we begin by applying an approximate Bayesian computation (ABC) procedure. In this phase, we simulate data under the model for each proposed parameter value and compute a distance to the observed data based on intensity histograms. A procedure similar to the one described above is then applied to generate 10,000 accepted samples. This initial sampling takes approximately four hours but is not reliable: the likelihood estimates for a fixed parameter value often vary by orders of magnitude, and the method tends to identify only a restricted region of parameter space where the observed fluorescence intensities are approximately matched. Nonetheless, we use the mean and standard deviation of these samples to define a Gaussian prior for the subsequent MCMC sampling described above.

\newpage
\section{Cleaning real dataset}\label{SIsec:SVM}

This section describes the data preprocessing pipeline applied to the flow cytometry measurements discussed in Sec. ~\ref{sec:real} and visualized in Figure~\ref{fig:real_results}a. The goal of this cleaning procedure is to isolate a consistent population of cells suitable for downstream analysis by removing artifacts and outliers that arise during acquisition or due to biological heterogeneity.

The dataset includes three channels of interest: {FSC-A} (forward scatter area), which relates to cell size; {FSC-H} (forward scatter height), which helps distinguish single cells from aggregates; and {FL1-A}, a fluorescence channel used to report gene expression.

As a first step, we removed events with clearly invalid measurements. Specifically, we excluded:
Events containing missing NaN values, and non-positive values in any of the key channels which correspond to debris or saturated detectors.
We also filtered out events where the ratio between FSC-A and FSC-H deviated by more than 30\% from 1. This ratio is expected to be close to unity for round, single cells, and deviations typically suggest doublets or irregular morphologies. 

To further refine the population, we applied a density-based outlier detection method aimed at identifying and removing events that lie outside the main cluster of the data. After log-transforming the FSC-A, FSC-H, and FL1-A measurements, we standardized the resulting feature space to remove correlations and ensure comparability between variables. Two empirical filters were also applied to focus the analysis on the densest region of the distribution, excluding events with extreme FSC-H/FSC-A ratios or low FSC-A values.

We then trained a one-class support vector machine (SVM) \cite{svm_scikit_learn} on a representative subset of this core population. The model was used to define a high-density region in feature space, effectively learning a boundary that separates the main population from peripheral or sparse regions. This approach enables consistent removal of rare or anomalous events across samples, without requiring manual gating.

\begin{figure}
    \centering
    \includegraphics[width=.9\textwidth]{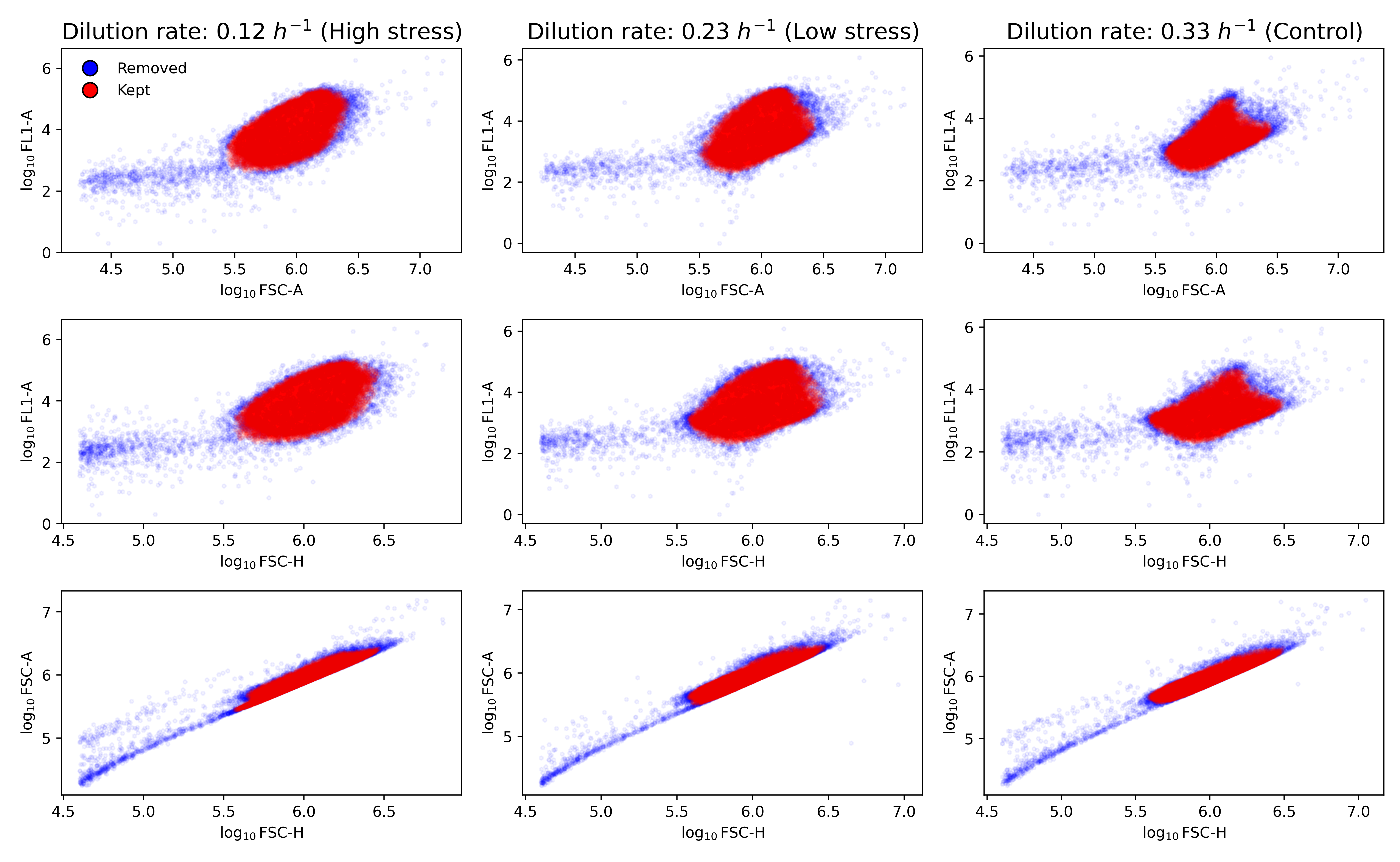}
    \caption{Outlier removal procedure applied to flow cytometry data across three dilution conditions. Each row corresponds to a different projection of the FSC-A, FSC-H, and FL1-A features. Blue points indicate events retained for analysis; red points were classified as outliers and removed. The procedure was applied independently to each condition, these are equivalent to the respective ``high stress'', ``low stress'', and ``control''  in Fig. \ref{fig:real_results}. }
    \label{fig:cleaning}
\end{figure}

This cleaning procedure was applied independently to each dilution condition to account for differences in cell density and distribution across experimental setups. The results of the filtering and outlier removal steps are illustrated in Fig. \ref{fig:cleaning}, where events removed by the algorithm are shown in red and retained events in blue. Each panel corresponds to a different pairwise projection of the log-transformed features used for classification, with each column coming from cells cultivated at the same dilution rate.

\newpage
\section{Why we assume active/inactive state}\label{SIsec:Real_data_active}
The model in Sec. \ref{sec:fc}, later applied in Sec. \ref{sec:real}, assumes that the cell switches between two states: an active state producing protein at rate $\beta$, and an inactive state with zero production.

While more complex models (e.g., multiple or non-parametric states) are conceivable, they must be reflected in the observable fluorescence data. Here we argue that it is reasonable to assume no production in the inactive state.
In the case where there is a single producing state, as described in Sec. \ref{sec:deterministic_cell_division}, the expected steady-state protein number $G$ in  \eqref{poisson_beta_marginalized} is given by $\langle G \rangle = \frac{3}{2} {\beta}{T}$, where $T$ is the cell division time. In chemostats, where division is set by the dilution rate $D$ (see SI \ref{SIsec:real_data_divisiontime}), this becomes $\langle G \rangle = \frac{3}{2} \frac{\beta}{D}$. If cells were always producing, slower dilution (and thus longer division times) would lead to higher average protein numbers and a shift in the fluorescence peak. 

Yet, in Fig. \ref{fig:real_results}a, the peak remains fixed across dilution rates, suggesting a true inactive (effectively zero-production) state compensates for longer division times.

\begin{figure}[h]
    \centering
    \includegraphics[width=0.9\textwidth]{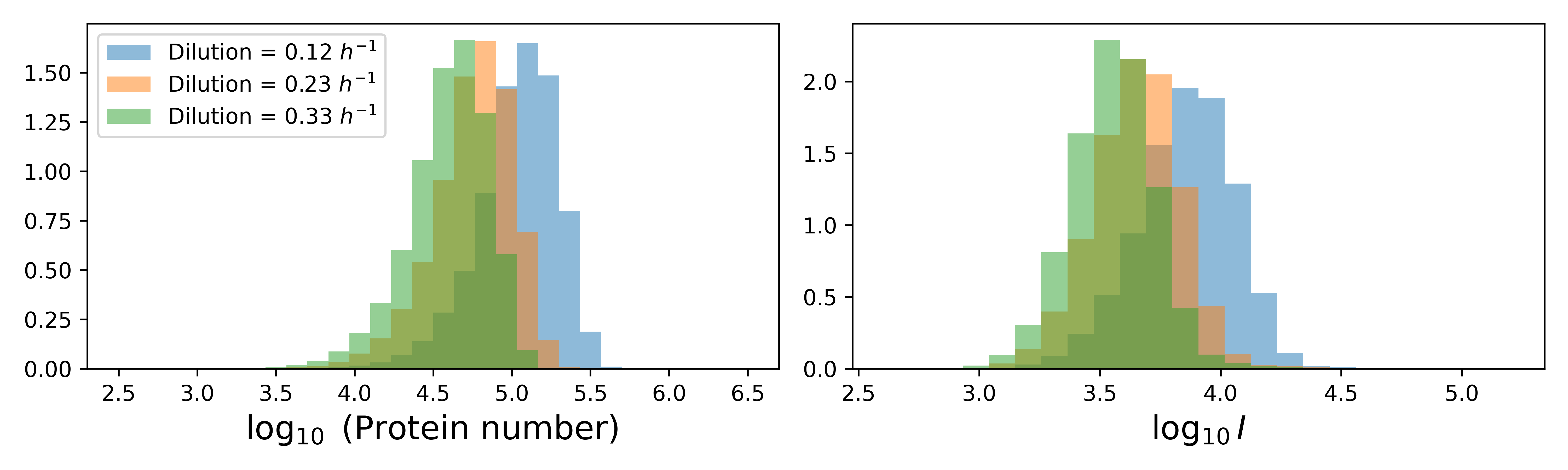}
    \caption{
        {Simulation of a one-state model with constant protein production}. 
        The protein number (left)  and resulting fluorescence distributions (right) for hypothetic case where we had the same autofluorescence distribution, but there were a contant  protein production rate of $10^4  h^{-1}$ (about two order of magnitude smaller than the results presented in Fig. \ref{fig:real_results}).
        As the dilution rate $D$ decreases, the peaks of the distributions shift to higher values due to longer accumulation times. 
        This behavior is inconsistent with the experimental data in Fig. \ref{fig:real_results}a, where peak the low stress has a peak . 
        The result supports the assumption of an inactive state with zero production in the real system.
    }
    \label{SI_onestate}
\end{figure}

This is supported by simulations presented in Fig. \ref{SI_onestate}, where a one-state always-producing model is used. There, protein synthesis on the order of $10^4 h^{-1}$  (two orders of magnitude lower than the levels inferred in Sec. \ref{sec:real})  leads to fluorescence distributions that change significantly with the dilution rate, even after accounting for autofluorescence. The discrepancy between these simulated peaks and the experimental data further supports the presence of a non-producing state.

 \newpage

\section{Pre-calibration of $\theta$ distribution for real data using ABC}
\label{SIsec:precalibration_real}

In Sec.  \ref{sec:real}, we mention that training the neural network on unrepresentative regions of parameter space can bias inference. To mitigate this, we initialize training using a prior informed by a simpler SBI method, Approximate Bayesian Computation (ABC). 
This method estimates a rough likelihood by comparing histograms of simulated flow cytometry intensity values at various parameters $\theta$ to the observed data shown in Fig. \ref{fig:real_results}. We then perform the MCMC sampling procedure detailed in Sec. \ref{SIsec:MCMC}. We use this MCMC sampled values to construct the prior for Sec. \ref{sec:real}. 
The specifics of the ABC approximation are provided in Sec. \ref{SIsec:abc_likelihood}, while Sec. \ref{SIsec:flow_comparison} explains why our normalizing flows-based approach yields a more reliable posterior.

\subsection{ABC-based likelihood}
\label{SIsec:abc_likelihood}

To construct the likelihood approximation using ABC, we discretize the observed log fluorescence intensity values, denoted $\log I$, into $A = 200$ equally spaced bins. These bins define intervals of the form $(L_a, L_{a+1})$, where
$ L_0 = \min(\log I_{\text{data}}), \quad L_A = \max(\log I_{\text{data}}), \quad \Delta L = L_{a+1} - L_a$.

Given a parameter vector $\theta$, we simulate $S = 2^{12}$ samples $I_s$ from the model $p(I | \theta)$, as described in Sec. \ref{sec:real}. The approximate likelihood of the log-intensity under these parameters is then defined as:
\begin{equation}
\label{eq:abc_likelihood}
p(\log I \mid \theta) \approx \frac{1}{\Delta L} \cdot \frac{1}{S + 1} \sum_{a=0}^{A-1} \mathbf{1}_{\log I \in (L_a, L_{a+1})} \left[ \frac{1}{A} + \sum_{s=1}^S \mathbf{1}_{\log I_s \in (L_a, L_{a+1})} \right],
\end{equation}
where $\mathbf{1}_{x \in (L_a, L_{a+1})}$ is an indicator function that equals 1 if $x$ falls within the interval $(L_a, L_{a+1})$, and 0 otherwise.
This likelihood estimate is proportional to the number of simulated values $\log I_s$ that fall into the same bin as the observed value, with a small pseudocount ($1/A$) added to each bin, avoiding zero likelihood.

Using this approximation, we proceed with the adaptive MCMC sampling scheme described in Sec.~\ref{SIsec:MCMC}. The mean and standard deviation of the resulting ABC posterior samples are used to define the prior distribution over parameters $\theta$.

Specifically, we assume a prior in which the logarithm of each parameter in $\theta$ follows a normal distribution. The means and standard deviations of this distribution, shown in Table~\ref{tab:res}, correspond to those of the ABC posterior. To ensure sufficient coverage of the posterior, standard deviations smaller than 0.5 are rounded up to the nearest half-integer. This prior also defines the distribution from which synthetic data are drawn during neural network training, as used in the results presented in Sec.~\ref{sec:real}.
\begin{table}
\centering
\begin{tabular}{ll|c|c}
\hline
$\quad$ & \textbf{Parameter} & \textbf{Prior Mean} & \textbf{Prior Std Dev} \\
\hline
 & $\ln \beta$ [$h^{-1}$] & 13.2 & 0.915 \\
 \hline
\multicolumn{2}{l|}{\textit{High stress} ($\overline{T} = 8.33 h$)} & \\
 & $\ln \lambda_{\text{act}}$ [$\overline{T}^{-1}$] & $-0.2$ & 0.5 \\
 & $\ln \lambda_{\text{ina}}$ [$\overline{T}^{-1}$] & $2.5$ & 1.192 \\
 & $\ln \sigma$  & $-1.4$ & 0.5 \\
\hline
\multicolumn{2}{l|}{\textit{Low stress} ($\overline{T}'  = 4.35 h$)} & \\
 & $\ln \lambda_{\text{act}}$ [$\overline{T}'^{-1}$] & $-2.2$ & 0.5 \\
 & $\ln \lambda_{\text{ina}}$ [$\overline{T}'^{-1}$] & $1.4$ & 1.132 \\
 & $\ln \sigma$  & $-2.8$ & 0.714 \\
\end{tabular}
\caption{Prior means and standard deviations for model parameters for the real data analysis in Sec.~\ref{sec:real}. Parameters for activation and inactivation rates ($\lambda_{\text{act}}$ and $\lambda_{\text{ina}}$) are expressed in units of inverse mean cell division time. Under high stress, $\overline{T} = 1/\gamma = 8.33 h$ with dilution rate $0.12 h^{-1}$. Under low stress, $\overline{T}' = 1/\gamma' = 4.35 h$ with $\gamma' = 0.23 h^{-1}$.}
\label{tab:res}
\end{table}

\subsection{Advantages of the normalizing flow-based posterior}
\label{SIsec:flow_comparison}

The normalizing flow posterior offers two principal advantages over the ABC-based likelihood approach.

\paragraph{1. Computational efficiency.} Once trained, the normalizing flow provides a closed-form, differentiable approximation of the posterior, enabling rapid sampling. For example, generating $10^4$ samples with the ABC-based MCMC scheme (Sec. \ref{sec:real}) required approximately 4 hours. In contrast, the same time budget with the normalizing flow–based posterior yielded 10 times more samples. This efficiency is especially beneficial when repeated inference or model refinement is required.

\paragraph{2. Improved MCMC mixing.} The normalizing flow posterior also enhances MCMC mixing behavior. 

To compare the behavior of the ABC-based and normalizing flows posteriors shown in Fig. \ref{fig:real_results}b, we trace the log-posterior values across MCMC iterations in both cases. In these plots, we subtract the maximum log-posterior value such that the maximum is zero, making relative differences more apparent. This transformation does not affect MCMC dynamics but highlights stability and convergence.

For the ABC-based posterior, we observe that after the initialization described in Sec.  \ref{SIsec:MCMC}, the chain does reach regions of higher posterior density. However, even after apparent convergence, the log-posterior continues to oscillate with a magnitude on the order of hundreds. Since the posterior is proportional to the exponential of the log-posterior, these fluctuations imply changes in probability mass by factors as large as $e^{100}$ an implausibly large variation in a supposedly stable region of parameter space. This indicates that the ABC approximation introduces significant noise, stemming from the coarse histogram-based likelihood estimate in \eqref{eq:abc_likelihood}.

In contrast, the normalizing flows–based posterior shows smoother convergence and stable sampling after burn-in, with log-posterior values remaining within a narrow band around the mode. This suggests that the flow provides a more accurate and numerically stable surrogate for the likelihood, yielding a better-behaved posterior landscape.

\begin{figure}
    \centering
    \vspace{-.5cm}
    \includegraphics[width=0.9\linewidth]{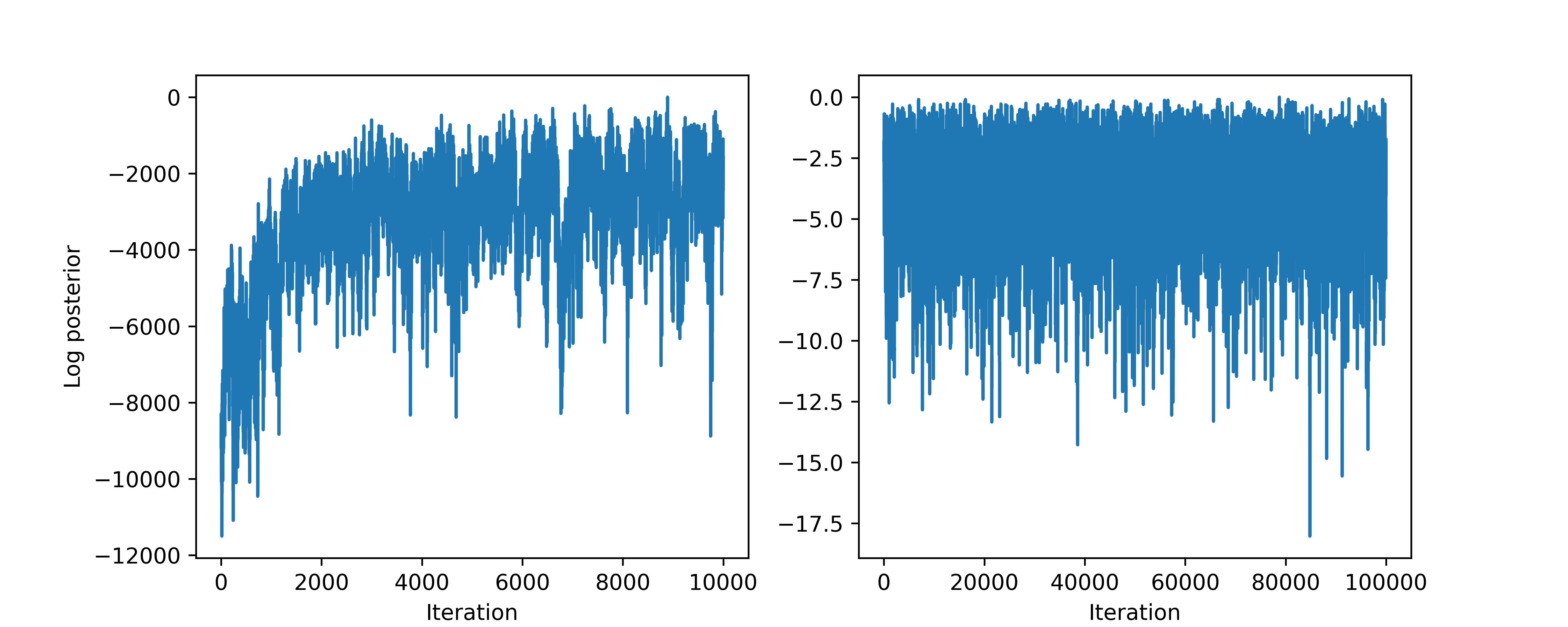}
    \vspace{-.5cm}
    \caption{Comparison of log-posterior traces across MCMC iterations for the ABC-based approximation (left) and the normalizing flow–based posterior (right). 
    Each trace has been shifted by subtracting the maximum log-posterior value to emphasize differences in convergence behavior.
    The ABC-based posterior shows large fluctuations even after burn-in, with log-posterior variations on the order of hundreds corresponding to probability changes by factors of $e^{100}$. This instability arises from the histogram-based likelihood approximation.
    In contrast, the normalizing flow–based posterior converges more cleanly and remains stable around the maximum, indicating better approximation of the true posterior landscape.}
    \label{fig:logposterior_traces}
\end{figure}

\end{document}